\documentclass[twocolumn,floatfix,showpacs,tightenlines,prb]{revtex4}
\usepackage{graphicx}
\begin{document}

\title{Quenching Bloch oscillations in a strongly correlated material} 
\author{J.~K.~Freericks}
\affiliation{Department of Physics, Georgetown University, Washington, DC
20057, U.S.A.}

\date{\today}

\begin{abstract}
Dynamical mean-field theory is generalized to solve the nonequilibrium Keldysh
boundary problem: a system is started in equilibrium at a temperature $T=0.1$, a
uniform electric field is turned on at $t=0$, and the system is monitored as it
approaches the steady state.  The focus here is on the Bloch oscillations of the current and
how they decay after their initial appearance near $t=0$.  The system is 
evolved out to the longest time allowed by our computational resources---in most
cases we are unable to reach the steady state.  The strongly correlated material
is modeled by the spinless Falicov-Kimball model at half-filling on a hypercubic
lattice in $d=\infty$ dimensions, which has a Mott-like metal-insulator 
transition at $U=\sqrt{2}$.  The computational algorithm employed is highly
efficient, parallelizes well, and scales to thousands of processors. For strong fields, we find beats develop with a period of $2\pi/U$, while for strong interactions, the Bloch oscillations are sharply damped and become quite irregular in time.
\end{abstract}

\pacs{71.27.+a, 71.10.Fd, 71.45.Gm, 72.20.Ht}

\maketitle

\section{Introduction}
Dynamical mean-field theory (DMFT)
was introduced\cite{brandt_mielsch_1989} in 1989 as a new technique
to solve the many-body problem.   Originally, DMFT was used to find the charge-density-wave transition
temperature of the spinless Falicov-Kimball model\cite{falicov_kimball_1969} at half-filling.
Since then, it has been employed to solve nearly all model solid-state Hamiltonians\cite{kotliar_review} in equilibrium.  It was recently generalized to treat nonequilibrium situations\cite{freericks_prl_2006}.  This contribution is a sequel to that work, where we examine the transient evolution of the Bloch oscillations of the current that appear in a system that starts in equilibrium and has a uniform electric field turned on at time $t=0$---the so-called Keldysh boundary problem\cite{keldysh_1965}. 

Bloch~\cite{bloch_1928} and Zener~\cite{zener_1934} showed that noninteracting electrons
(on a lattice) undergo an oscillatory motion when placed in a uniform
static electric field (oriented along a symmetry direction of the Brillouin zone), because the electron wavevector, which evolves linearly in time under
the driving force from the electric field, is Bragg reflected whenever it reaches a Brillouin zone 
boundary (since the wavevector will periodically trace out an identical orbit when the electric field is oriented in a symmetry direction). But in conventional metals, Bloch oscillations have never been seen, because
the electron relaxation time is so short, the electrons are scattered 
before they reach the zone boundary and
Bragg reflect.  Bloch oscillations have been observed in semiconducting
heterostructures~\cite{bloch_semi}, Josephson junctions~\cite{bloch_josephson},
and cold-atom systems~\cite{bloch_atom}.   In this contribution, we use nonequilibrium DMFT to solve the Keldysh boundary problem and analyze how the Bloch oscillations are quenched in a strongly correlated material as the scattering is increased.  In particular, we describe the change in the behavior of the current as the system evolves from a metal to a Mott insulator with increasing $U$.

The formalism for the nonequilibrium many-body problem was developed in the 1960s by Kadanoff and Baym\cite{kadanoff_baym_1962} and Keldysh\cite{keldysh_1965}.  They determined the generic equations for the Green's functions and then solved them under different approximate conditions. Since then, the exact solution has remained one of the long-standing unsolved problems of many-body physics.  Now, DMFT can be employed to solve this problem in the limit of large spatial dimensions. In this contribution, we describe, in detail, how to do this.  This problem is quite complex, and necessarily, we need to limit our coverage to focus the discussion.  We have chosen to concentrate on the phenomena of the quenching of Bloch oscillations, examining how they change their character as one goes from a metal to a Mott insulator, and how they develop beats for large fields.

The remainder of this contribution is organized as follows: in Section II, we discuss the formalism, and derive, in detail, all of the relevant generalizations of DMFT to treat the nonequilibrium case as described by the Keldysh boundary problem.  In Section III, we present our numerical results for extensive calculations of the current as a function of time on the hypercubic lattice. We will see a range of interesting behavior in both the high field limit and in the Mott insulator. We end with a discussion of the results and a conclusion in Section IV.

\section{Formalism}
\begin{figure}[h]
\centering{
\includegraphics[width=1.5in,angle=0]{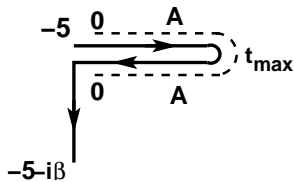}}
\caption{
Kadanoff-Baym-Keldysh contour for the two-time Green's functions
in the nonequilibrium case. We take the contour to run from $t_{min}=-5$ to
$t_{max}$ and back, and then extend downward parallel to the imaginary
axis for a distance of $\beta$.  The field is turned on at $t=0$;
{\it i.e.}, the vector potential is nonzero only for positive times.
}
\label{fig: contour}
\end{figure}

The many-body formalism for nonequilibrium dynamical
mean-field theory is straightforward to develop within the Kadanoff-Baym-Keldysh
approach~\cite{kadanoff_baym_1962,keldysh_1965}.
Because nonequilibrium problems are not time-translation invariant (indeed an external field is turned on at $t=0$), we need
to employ Green's functions that depend on two times.  We work with the so-called
contour-ordered Green's function, which is defined for 
any two time values that lie on the Kadanoff-Baym-Keldysh contour shown in
Fig.~\ref{fig: contour}.  The system starts in equilibrium until
time $t=0$ when a uniform electric field is turned on.  The contour starts at some time
before the field is turned on, runs out to a maximal time, then returns to
the original time, and finally moves parallel to the negative imaginary
axis a distance $\beta$ (equal to the inverse of the temperature of the
original equilibrium distribution). 
The evolution forward, and then backward, 
in time is necessary because there is no {\it a priori} relationship
between the quantum states at large positive times, and the original
equilibrium distribution, so we must evolve backward in time to reach a
known equilibrium distribution.

We will be working with a spatially uniform electric field.  We describe it via a spatially
uniform vector potential (with a vanishing scalar potential) in the so-called Hamiltonian
gauge (see below).  In this case, we preserve the translational invariance of the system 
in position space, so we can describe results in either real space or momentum space.
Working in the Heisenberg picture, where all time dependence is in the operators,
the contour-ordered Green's function (in real space) is defined by
\begin{eqnarray}
G^c_{ij}(t,t^\prime)&=&-i{\rm Tr} \mathcal{T}_c 
e^{-\beta \mathcal{H}_{eq}}c_i^{}(t)c^\dagger_j(t^\prime)/\mathcal{Z}_{eq},
\label{eq: g_contour_def}\\
&=&
-i\theta_c(t,t^\prime){\rm Tr}e^{-\beta \mathcal{H}_{eq}}c_i^{}(t)c^\dagger_j(t^\prime)/\mathcal{Z}_{eq}\nonumber\\
&~&+i\theta_c(t^\prime,t){\rm Tr} e^{-\beta\mathcal{H}_{eq}}c^\dagger_j(t^\prime)c_i^{}(t)/\mathcal{Z}_{eq},
\label{eq: h_contour_def2}
\end{eqnarray}
where $c_i^\dagger$ ($c_i^{}$) are the electron creation (annihilation) operators for conduction electrons at site $i$, the subscript $eq$ denotes the equilibrium Hamiltonian
(chosen at any time prior to when the field is turned on), and the equilibrium partition function satisfies $\mathcal{Z}_{eq}={\rm Tr}\exp[-\beta \mathcal{H}_{eq}]$, with $\beta$
the inverse temperature. The symbol $\theta_c(t,t^\prime)$ is the generalization of the theta function to the contour, and it equals 1 if $t$ lies after $t^\prime$ on the contour and it equals 0 otherwise. The time-dependence of the fermionic operators is expressed in the Heisenberg representation with the time-evolution taking place along the contour (the times $t$ and $t^\prime$ are any two times on the three-branch contour).

By choosing the time variables to lie on different pieces of the contour, we can determine different types of Green's functions\cite{wagner_1991}.  In particular, if both $t$ and $t^\prime$ lie on the upper real branch, we have the time-ordered Green's function
\begin{equation}
 G^t_{ij}(t,t^\prime)=-i{\rm Tr} \mathcal{T}_t
e^{-\beta \mathcal{H}_{eq}}c_i^{}(t)c^\dagger_j(t^\prime)/\mathcal{Z}_{eq},
\label{eq: g_timeordered_def}
\end{equation}
if both $t$ and $t^\prime$ are on the lower real branch, we have the anti-time-ordered
Green's function
\begin{equation}
 G^{\bar t}_{ij}(t,t^\prime)=-i{\rm Tr} \bar\mathcal{T}_t
e^{-\beta \mathcal{H}_{eq}}c_i^{}(t)c^\dagger_j(t^\prime)/\mathcal{Z}_{eq},
\label{eq: g_antitimeordered_def}
\end{equation}
where the bar denotes anti-time-ordering.  Similarly, when $t$ lies on the upper real branch
and $t^\prime$ lies on the lower real branch, we have the so-called lesser Green's
function
\begin{equation}
 G^<_{ij}(t,t^\prime)=i{\rm Tr} 
e^{-\beta \mathcal{H}_{eq}}c^\dagger_j(t^\prime)c_i^{}(t)/\mathcal{Z}_{eq},
\label{eq: g_lesser_def}
\end{equation}
and when $t$ lies on the lower real branch and $t^\prime$ lies on the upper real branch, we have
the so-called greater Green's function
\begin{equation}
 G^>_{ij}(t,t^\prime)=-i{\rm Tr} 
e^{-\beta \mathcal{H}_{eq}}c_i^{}(t)c^\dagger_j(t^\prime)/\mathcal{Z}_{eq}.
\label{eq: g_greater_def}
\end{equation}
Finally, when both times lie on the imaginary branch, we have the thermal Green's function
(because the Hamiltonian is the equilibrium Hamiltonian here). There also are mixed Green's
functions, where one time argument lies on a real branch and the other on the imaginary branch.  These Green's functions are required to properly determine the transient response at short times. They also enter when we map the lattice problem onto the impurity problem in an additional time-dependent field.  Note that we can similarly define corresponding momentum-dependent Green's functions in analogy to the real-space Green's functions defined above.

In addition to the Green's functions that can be directly extracted from the contour-ordered Green's function, there are two other important Green's functions that can be defined.  The retarded Green's function is defined for real times, and satisfies
\begin{equation}
 G^R_{ij}(t,t^\prime)=-i\theta(t-t^\prime){\rm Tr}e^{-\beta \mathcal{H}_{eq}}
\{c_i^{}(t),c^\dagger_j(t^\prime)\}_+/\mathcal{Z}_{eq},
\label{eq: g_retarded_def}
\end{equation}
and the advanced Green's function satisfies
\begin{equation}
 G^A_{ij}(t,t^\prime)=i\theta(t^\prime-t){\rm Tr}e^{-\beta \mathcal{H}_{eq}}
\{c_i^{}(t),c^\dagger_j(t^\prime)\}_+/\mathcal{Z}_{eq},
\label{eq: g_advanced_def}
\end{equation}
where the curly brackets denote the anticommutator.

In nonequilibrium physics problems, we focus primarily on two of these different Green's functions---the retarded Green's function, which determines the quantum-mechanical states of the system, and the lesser Green's function, which determines how the electrons are distributed amongst those quantum states. It turns out that all of the other Green's functions defined with real time arguments can be constructed from these two ``fundamental'' Green's functions.

We will consider a uniform electric field turned on at time $t=0$. We ignore all magnetic 
field and relativistic effects and assume the field is always uniform in space.  Then we can describe the field via a spatially uniform time-dependent vector potential in the Hamiltonian gauge, where the scalar potential vanishes
\begin{equation}
 {\bf E}=-\frac{d}{dt}{\bf A}(t);
\label{eq: hamiltonian_gauge}
\end{equation}
here we have set the speed of light equal to one. The noninteracting contour-ordered Green's function in momentum space is
\begin{equation}
 G^{c,non}_{\bf k}(t,t^\prime)=-i{\rm Tr}\mathcal{T}_c e^{-\beta\mathcal{H}_{eq,non}}
c_{\bf k}^{}(t)c_{\bf k}^\dagger(t^\prime)/\mathcal{Z}_{eq,non},
\label{eq: g_contour_non}
\end{equation}
where we use creation and annihilation operators in momentum space, and evaluate all averages with respect to the equilibrium noninteracting Hamiltonian
\begin{equation}
 \mathcal{H}_{eq,non}=\sum_{\bf k}(\epsilon_{\bf k}-\mu)c^\dagger_{\bf k}c^{}_{\bf k},
\label{eq: ham_eq_non}
\end{equation}
with the bandstructure satisfying $\epsilon_{\bf k}=\lim_{d\rightarrow\infty}-t^*\sum_{i=1}^d\cos({\bf k}_i)/\sqrt{d}$.
The nearest neighbor hopping is chosen to scale\cite{metzner_vollhardt_1989} as $t=t^*/2\sqrt{d}$,
and we use $t^*$ as our energy unit (the lattice constant $a$ is also set equal to 1). The electric field is introduced via the vector potential, which is chosen to be
\begin{equation}
 {\bf A}(t)=-\theta(t){\bf E}t,
\label{eq: vector_potential}
\end{equation}
for a uniform field turned on at $t=0$. The time-dependent Hamiltonian is then constructed via the Peierls' substitution\cite{peierls}, where ${\bf k}\rightarrow {\bf k}-e{\bf A}(t)$ in the bandstructure.  Hence the time-dependent noninteracting Hamiltonian becomes
\begin{equation}
 \mathcal{H}_{non}=\sum_{\bf k}[\epsilon_{{\bf k}+\theta(t)e{\bf E}t}-\mu]
c^\dagger_{\bf k}c^{}_{\bf k}.
\label{eq: ham_non}
\end{equation}
Note that the noninteracting Hamiltonian commutes with itself at all times, even though it is time-dependent.  This result makes it easy to exactly solve for the noninteracting Green's functions in a field\cite{jauho_wilkins,turkowski_freericks_2005}.  First note that the time-dependent creation and annihilation operators satisfy
\begin{eqnarray}
 c^\dagger_{\bf k}(t)&=&\exp\left ( i\int_{t_{min}}^t d\bar t \left [ \epsilon_{{\bf k}+\theta(\bar t)e{\bf E}\bar t}-\mu\right ]\right ) c_{\bf k}^\dagger,
\label{eq: c_dagger_t}\\
c^{}_{\bf k}(t)&=&\exp\left ( -i\int_{t_{min}}^t d\bar t \left [ \epsilon_{{\bf k}+\theta(\bar t)e{\bf E}\bar t}-\mu\right ]\right ) c_{\bf k}^{},
\label{eq: c_t}
\end{eqnarray}
which follows by directly integrating their equations of motion (the integration in time runs \textit{over the contour} from the starting point at $t_{min}$ to the desired time $t$ because we need the time arguments for any location on the contour; $t_{min}=-5$ for the calculations presented here). Substituting into the definition for the noninteracting contour-ordered Green's function in Eq.~(\ref{eq: g_contour_non}) then produces the noninteracting Green's function directly (the integral over time lies on the contour between the points $t^\prime$ and $t$)
\begin{eqnarray}
 G^{c,non}_{\bf k}(t,t^\prime)&=&i[f(\epsilon_{\bf k}-\mu)-\theta_c(t,t^\prime)]e^{i\mu(t-t^\prime)}\nonumber\\
&\times&\exp\left ( -i\int^t_{t^\prime}d\bar t \epsilon_{{\bf k}+\theta(\bar t)e{\bf E}\bar t}\right ).
\label{eq: g_contour_non_final}
\end{eqnarray}
The Fermi-Dirac distribution $f(\epsilon_{\bf k}-\mu)=1/[1+\exp(\beta\{\epsilon_{\bf k}-\mu\})]$ enters from the initial equilibrium distribution of the electrons (prior to the field being turned on).

The noninteracting Green's function is quite complicated for an electric field pointing in an arbitrary direction.  It turns out that the simplest problem to consider is one where the field points along a diagonal direction ${\bf E}=E(1,1,1,...,1)$.  Then the noninteracting Green's function depends on two band energies: $\epsilon_{\bf k}$ (already defined above) and 
$\bar\epsilon_{\bf k}=\lim_{d\rightarrow\infty}-t^*\sum_{i=1}^d\sin{\bf k}_i/\sqrt{d}$ in the following way:
\begin{eqnarray}
 G^{c,non}_{\epsilon,\bar\epsilon}(t,t^\prime)&=&i[f(\epsilon-\mu)-\theta_c(t,t^\prime)]e^{i\mu(t-t^\prime)}\nonumber\\
&\times&\exp\Biggr ( -i\int^t_{t^\prime}d\bar t \{ [\theta(-\bar t)+\theta(\bar t)\cos(eE\bar t)]\epsilon \nonumber\\
&~&- \theta(\bar t)\bar\epsilon\sin(eE\bar t)\}\Biggr ).
\label{eq: g_contour_non_final2}
\end{eqnarray}
The two band energies are distributed with a joint density of states that is a Gaussian in each variable\cite{turkowski_freericks_2005}
\begin{equation}
 \rho(\epsilon,\bar\epsilon)=\frac{1}{\pi}\exp(-\epsilon^2-\bar\epsilon^2).
\label{eq: joint_dos}
\end{equation}
It will turn out that all momentum-dependent quantities we are interested in will depend only on $\epsilon$ and $\bar\epsilon$, so we reduce the infinite-dimensional Brillouin zone to a two-dimensional energy space.  Note that in equilibrium, all $\bar\epsilon$ dependence drops out, and we have all objects of interest depending only on the bandstructure $\epsilon$.

The contour-ordered Green's function for the interacting many-body problem satisfies Dyson's equation (with all time integrals running over the contour)
\begin{eqnarray}
 G^c_{\bf k}(t,t^\prime)&=&G^{c,non}_{\bf k}(t,t^\prime)\label{eq: dyson}\\
&+&\int_c d\bar t \int_c d\bar t^\prime G^{c,non}_{\bf k}(t,\bar t)\Sigma^c(\bar t,\bar t^\prime)G^c_{\bf k}(\bar t^\prime,t^\prime),\nonumber
\end{eqnarray}
which can be viewed as defining the contour-ordered self-energy.  We have written out the Dyson equation explicitly in momentum space, and we used the fact that the self-energy is local (and hence independent of momentum) in DMFT, a fact that is proven next.

The argument that the self-energy is local in $d\rightarrow \infty$ is based on two facts---the expansion of the self-energy in powers of the hopping\cite{metzner_1991} which show that the self-energy is local (in equilibrium) and the Langreth rules\cite{langreth} which show how to relate perturbation theory in equilibrium to perturbation theory for the nonequilibrium case, and thereby shows that the nonequilibrium self-energy is also local. This follows because the many-body perturbation theory diagrams are identical in structure
for both the equilibrium and nonequilibrium perturbation theories,
so the power-counting analysis (in inverse powers of the spatial dimension $d$) of Metzner guarantees that
the nonequilibrium self-energy remains local in DMFT.  In other words, the nonequilibrium
DMFT problem can be mapped onto an impurity problem in time-dependent
fields, just like the equilibrium problem, \textit{except now the fields (the Green's functions and the self-energy) have two time arguments}. 

The DMFT algorithm employed to solve for the transient response of nonequilibrium systems is closely related to the iterative approach used in equilibrium\cite{jarrell_1992}. We develop each of the steps of this algorithm next, and then describe the differences between the nonequilibrium calculations and the equilibrium ones.

The first step is to generalize the Hilbert transform employed in equilibrium DMFT
to the nonequilibrium case. Starting from the local self-energy $\Sigma^c(t,t^\prime)$,
which is a continuous matrix operator with both time variables running over the contour, we need to construct the local contour-ordered Green's function $G^c_{loc}(t,t^\prime)=\sum_{\bf k}G^c_{\bf k}(t,t^\prime)$.  This is done by first expressing the momentum dependence in terms of the two band energies $\epsilon$ and $\bar\epsilon$, and employing the Dyson equation in Eq.~(\ref{eq: dyson}).  Hence,
\begin{eqnarray}
 G^c_{loc}(t,t^\prime)&=&\int d\epsilon \int d\bar\epsilon \rho(\epsilon,\bar\epsilon)\nonumber\\
&\times&\left [ \left ( I-G^{c,non}_{\epsilon,\bar\epsilon}\Sigma^c\right )^{-1} G^{c,non}_{\epsilon,\bar\epsilon}\right ](t,t^\prime),
\label{eq: g_loc}
\end{eqnarray}
where $I$ denotes the identity matrix operator [equal to $\delta_c(t,t^\prime)$---the Dirac delta function on the contour, which satisfies $\int_c dt^\prime \delta_c(t,t^\prime)f(t^\prime)=f(t)$], $G^{c,non}_{\epsilon,\bar\epsilon}$ is given in Eq.~(\ref{eq: g_contour_non_final2}), and there are a number of implicit matrix multiplications over the time variables (which range over the contour) on the right hand side. The integrand for each $\epsilon$ and $\bar \epsilon$ point is found by performing two continuous matrix operator multiplications and one continuous matrix operator inversion. Hence, the Hilbert transform for equilibrium problems generalizes to a two-dimensional integral of a matrix valued function that requires two matrix multiplies and one matrix inversion to determine the integrand.

Once the local Green's function has been determined, the next step is to extract the dynamical mean-field $\lambda^c(t,t^\prime)$.  This is determined by first finding the effective medium by using Dyson's equation and then extracting the dynamical mean field from the effective medium.  In particular, we have the effective medium satisfying
\begin{equation}
 G^c_0(t,t^\prime)=\left [ (G^c_{loc})^{-1}+\Sigma^c\right ]^{-1}(t,t^\prime),
\label{eq: effective_medium}
\end{equation}
and then the dynamical mean field becomes
\begin{eqnarray}
 \lambda^c(t,t^\prime)&=&(i\partial_t^c+\mu) \delta_c(t,t^\prime)-(G^c_0)^{-1}(t,t^\prime)\label{eq: dynamical_mean_field}\\
&=&(i\partial_t^c+\mu) \delta_c(t,t^\prime)-(G^c_{loc})^{-1}(t,t^\prime)+\Sigma^c(t,t^\prime).
\nonumber
\end{eqnarray}
Here $\partial^c_t$ is the derivative operator along the contour, so it equals $+\partial/\partial t$ along the upper branch, $-\partial/\partial t$ on the lower branch, and $-i\partial/\partial\tau$ along the negative imaginary time axis. Note that we calculate $\lambda^c$ from $[G_0^c]^{-1}$, so we do not normally compute $G_0^c$ to find the dynamical mean field.

Once the dynamical mean field has been found, we need to determine the impurity Green's function for the impurity problem that evolves in the presence of the dynamical mean field (which is nonzero on the entire contour). For the Falicov-Kimball model one can immediately write down the solution to this problem (at the moment, we do not have any numerical techniques to solve this problem for other models in the time representation).  The spinless Falicov-Kimball model~\cite{falicov_kimball_1969} involves single-band conduction electrons hopping on a lattice, and localized electrons which do not move
but do interact with the conduction electrons when they are in the same
unit cell via a screened Coulomb interaction $U$.  The equilibrium lattice Hamiltonian (in the
absence of a field) is then
\begin{equation}
\mathcal{H}_{eq}=-\frac{t^*}{2\sqrt{d}}\sum_{\langle ij\rangle}
(c^\dagger_ic^{}_j+c^\dagger_jc^{}_i)-\mu\sum_ic^\dagger_ic^{}_i+U\sum_i w_ic^\dagger_ic^{}_i.
\label{eq: hamiltonian_fk}
\end{equation}
Here, we have $c^\dagger_i$ ($c^{}_i$) create (annihilate) a spinless conduction
electron at site $i$, $w_i=0$ or 1 is the localized electron number operator
at site $i$, and $\mu$ is the conduction electron chemical potential.  Although the Falicov-Kimball model is one of the simplest many-body physics
models, it does have a Mott metal-insulator transition,
so it is interesting to use this model to examine how Bloch oscillations are quenched due to strong electron correlations 
(but the model does not include any Zener tunneling effects because there are 
no higher energy bands). The solution to the impurity problem can be found by 
solving the equations of motion for the contour-ordered Green's function (the procedure is essentially the same as in equilibrium except the Green's functions depend on two times now and the times run over the contour)
resulting in
\begin{eqnarray}
G_{imp}(t,t^\prime)&=&(1-\langle w_1\rangle)
\left [ (i\partial^c_t+\mu)\delta_c(t,t^\prime)-\lambda(t,t^\prime)
\right ]^{-1}\nonumber\\
&+&\langle w_1\rangle
\left [ (i\partial^c_t+\mu-U)\delta_c(t,t^\prime)
-\lambda(t,t^\prime)\right ]^{-1},\nonumber\\
\label{eq: g_imp_solve}
\end{eqnarray}
with $\langle w_1\rangle=\sum_i\langle w_i\rangle/N$ 
the average localized electron filling.

The Dyson equation in
Eq.~(\ref{eq: effective_medium}) is then employed to extract the impurity
self-energy from the impurity Green's function and the effective medium. The algorithm is then iterated until it converges (we usually require the Green's functions to converge pointwise to better than one part in $10^6$ in order to end the calculation).

In summary, the basic structure of the iterative approach to solving the
DMFT equations~\cite{jarrell_1992} continues to hold.  We start with a guess
for the self-energy (which is usually chosen to be equal to the 
equilibrium self-energy), then
we sum the momentum-dependent Green's function over the Brillouin zone to 
produce the local Green's function. Next the dynamical mean-field for
the impurity problem is extracted by using Dyson's equation for the 
local Green's function and self-energy, the impurity problem is solved in the
dynamical mean-field to produce the impurity Green's function, and Dyson's
equation is used again to extract the impurity self-energy.  In the 
self-consistent solution of the DMFT equations, the impurity self-energy will
be equal to the lattice self-energy.  If they are different, then the new
lattice self-energy is taken to be equal to the new impurity self-energy,
and the loop is iterated until it converges.  The nonequilibrium algorithm
includes the following modifications from the equilibrium algorithm: (i) the
summation over the Brillouin zone now requires at least a double integral
over two band energies; (ii) the Green's functions are described by continuous matrix operators with time indices that run over the contour; and (iii) the impurity
problem solver must be generalized to the nonequilibrium case.

The objects we work with on the contour are continuous matrix operators.  Unfortunately, there is no way to directly calculate with such objects on a computer.  Instead, we need to first discretize the contour, and define the continuous matrix operators as the limit where the discretization size goes to zero of discrete matrices, whose matrix elements are identified with the average value of the matrix operators within the corresponding discrete intervals on the contour.  By performing calculations with particular discretizations, and then taking the limit where the discretization goes to zero (via an extrapolation procedure), we can approximate the results for the continuous matrix operators. The contour is discretized in the following manner: we choose a real-time
spacing $\Delta t$ which varies from 0.1 to 0.014, and we fix the spacing
along the imaginary axis to $\Delta \tau=0.1i$ so our largest matrix
is $5700\times 5700$ and we evaluate integrals over the 
contour by discrete summations using the leftpoint rectangular
integration rule.  The matrix operators are general complex matrices,
which are manipulated using standard linear algebra packages (LAPACK and BLAS).

More precisely, the discretization process involves $N_t$ points on the upper real branch (ranging from $t_{min}$ to $t_{max}-\Delta t$), $N_t$ points on the lower real branch (ranging from $t_{max}$ to $t_{min}+\Delta t$), and 100 points along the imaginary axis (ranging from $t_{min}$ to $t_{min}-i\beta+0.1i$, with $\beta=10$); hence $\Delta t=(t_{max}-t_{min})/N_t$. The discrete time values on the contour are then
\begin{eqnarray}
 t_j&=&-t_{min}+(j-1)\Delta t,\quad\quad\quad 1\le j\le N_t,\label{eq: time_discrete}\\
&=&t_{max}-(j-N_t-1)\Delta t,\quad  N_t+1\le j\le 2N_t,\nonumber\\
&=&t_{min}-0.1i(j-2N_t-1),~ 2N_t+1\le j\le 2N_t+100,
\nonumber
\end{eqnarray}
where we used the fact that the discretization along the imaginary axis is always fixed at $\Delta\tau=0.1$ in our calculations.  To calculate integrals over the contour, we use a leftpoint rectangular integration rule for discretizing integrals. 
\begin{equation}
 \int_c dt f(t)=\sum_{i=1}^{2N_t+100}W_if(t_i),
\label{eq: integration_rule}
\end{equation}
where the weights satisfy
\begin{eqnarray}
 W_j&=&\Delta t,\quad\quad\quad 1\le j\le N_t,\nonumber\\
&=&-\Delta t,\quad\quad  N_t+1\le j\le 2N_t,\nonumber\\
&=&-0.1i,\quad\quad 2N_t+1\le j\le 2N_t+100.
\label{eq: integration_weights}
\end{eqnarray}
The leftpoint integration rule evaluates the function at the ``earliest'' point in the time interval that has been discretized for the quadrature rule (which is the left hand side of the interval when we are on the upper real branch; earliest is meant with regards to the sense that the contour is traversed).

Note that the contour-ordered Green's function satisfies a boundary condition where we identify the points $t_{min}$ with $t_{min}-i\beta$. One can show from the definition of the contour-ordered Green's function in Eq.~(\ref{eq: g_contour_def}), and the invariance of the trace with respect to the ordering of operators that $G^c_{ii}(t_{min},t^\prime)=-G^c_{ii}(t_{min}-i\beta,t^\prime)$
and $G^c_{ii}(t,t_{min})=-G^c_{ii}(t,t_{min}-i\beta)$.  This is similar to the antiperiodicity property that the thermal Green's functions satisfy, and the proof is identical.

The delta function changes sign along the negative real time branch, and
is imaginary along the last branch of the contour in order to satisfy the property that $\int_c dt^\prime \delta_c(t,t^\prime)f(t^\prime)=f(t)$.  In addition, we find that the numerics work better if the definition of the delta function is done via ``point splitting''  (when we calculate the inverse of a Green's function) so that the delta function does not lie on the diagonal, but rather on the first subdiagonal matrix (in the limit as $\Delta t\rightarrow 0$ it becomes a diagonal operator). Because we identify the times $t_{min}$ and $t_{min}-i\beta$, the point splitting approach to the definition of the delta function allows us to incorporate the correct boundary condition into the definition of the discretized delta function. Hence, we define the discretized delta function in terms of the quadrature weights, in the following way
\begin{eqnarray}
 \delta_c(t_i,t_j)&=&\frac{1}{W_i}\delta_{ij+1}, \quad{\rm for~integration~over~}j,\label{eq: delta_cont_defa}\\
&=&\frac{1}{W_{i-1}}\delta_{ij+1},~{\rm for~integration~over~}i,
\label{eq: delta_cont_defb}
\end{eqnarray}
where $t_i$ and $t_j$ are two points on the discretized contour as described in Eq.~(\ref{eq: time_discrete}), and $W_i$ are the quadrature weights described in Eq.~(\ref{eq: integration_weights}). We have a different formula for integration over the first variable versus integration over the second variable because we are using the leftpoint quadrature rule. Note that the formulas in Eqs.~(\ref{eq: delta_cont_defa}) and (\ref{eq: delta_cont_defb}) hold only when $i\ne 1$. When $i=1$, the only nonzero matrix element for the discretized delta function is the  $1,j=2N_t+100$ matrix element, and it has a sign change due to the boundary condition that the Green's function satisfies.
The discretization of the derivative of the delta function on the contour is even more complicated. It is needed to determine the inverse of the effective medium operator for the impurity.  The derivative is calculated by a two-point discretization that involves the diagonal and the first subdiagonal.  Since all we need is the discrete representation of the operator $[i\partial^c_t+\mu]\delta_c(t,t^\prime)$, we summarize the discretization of that operator as
follows
\begin{equation}
 [i\partial_t+\mu]\delta_c(t_j,t_k)=i\frac{1}{W_j}M_{jk}\frac{1}{W_k},
\end{equation}
\begin{widetext}
with the matrix $M_{jk}$ satisfying
\begin{equation}
 M_{jk}=\left (
\begin{array}{c c c c c c c c c c c c c}
 1 & 0 & 0 & & & & &...& & & & & 1+i\Delta t \mu\\
-1-i\Delta t \mu & 1 & 0 & & & & & ... & & & & & 0\\
0 & -1-i\Delta t \mu & 1 & 0 \\
 & & & \ddots \\
& & & 0 & -1+i\Delta t\mu & 1 & 0\\
& & & & 0 & -1+i\Delta t\mu & 1\\
& & & & & & &\ddots\\
& & & & & & & & -1-\Delta\tau\mu & 1\\
& & & & & & & & & & \ddots\\
& & & & & & & & & & & -1-\Delta\tau\mu& 1
\end{array}
\right );
\end{equation}
\end{widetext}
here $\Delta\tau=0.1$. The top third of the matrix corresponds to the upper real branch, the middle third to the lower real branch and the bottom third to the imaginary branch.
Note that the operator $[i\partial_t^c+\mu]\delta_c$ is the inverse operator of the Green's function of a spinless electron with a chemical potential $\mu$.  Hence the determinant of this operator must equal the partition function of a spinless electron in a chemical potential $\mu$, namely $1+\exp[\beta\mu]$. Taking the determinant of the matrix $M_{jk}$ gives
\begin{eqnarray}
 \det M&=&1 +(-1)^{2N_t+N_\tau}(1+i\Delta t\mu)(-1-i\Delta t\mu)^{N_t-1}\nonumber\\
&\times&(-1+i\Delta t\mu)^{N_t}(-1-\Delta\tau\mu)^{N_\tau},\nonumber\\
&\approx&1+(1+\Delta\tau\mu)^{N_\tau}+O(\Delta t^2),
\label{eq: det}
\end{eqnarray}
which becomes $1+\exp[\beta\mu]$ in the limit where $\Delta t,\Delta\tau\rightarrow 0$ ($N_\tau$ is the number of discretization points on the imaginary axis).  This provides a check on our algebra, and shows the importance of the upper right hand matrix element of the operator, which is required to produce the correct boundary condition for the Green's function.  This is also the reason why we chose to point-split the delta function when we defined it's discretized matrix operator.

We also have to show how to discretize the continuous matrix operator multiplication and how to find the discretized approximation to the continuous matrix operator inverse.  Matrix multiplication is discretized as follows
\begin{equation}
 \int_c d\bar t A(t,\bar t)B(\bar t,t^\prime)=\sum_k A(t_i,t_k)W_kB(t_k,t_j).
\label{eq; matrix_multiply}
\end{equation}
So we must multiply the columns (or the rows) of the discrete matrix by the corresponding quadrature weight factors.  This can be done either to the matrix on the left (columns) or to the matrix on the right (rows). To calculate the inverse, we recall the definition of the inverse for the continuous matrix operator
\begin{equation}
 \int_c d\bar t A(t,\bar t)A^{-1}(\bar t, t^\prime)=\delta_c(t,t^\prime),
\label{eq: matrix_inverse}
\end{equation}
which discretizes to
\begin{equation}
\sum_k A(t_i,t_k)W_kA^{-1}(t_k,t_j)=\frac{1}{W_i}\delta_{ij},
\label{eq: matrix_inverse_discrete}
\end{equation}
where here we do not need to point-split the delta function.  Hence, the inverse of the matrix is found by inverting the matrix $W_iA(t_i,t_j)W_j$, or, in other words, we must multiply the rows and the columns by the quadrature weights before using conventional linear algebra inversion routines to find the discretized version of the continuous matrix operator inverse.

Having resolved the technical details of the discretization, which allows us to evaluate approximations to the continuous matrix operators on a computer, we now move to the last technical aspect, which is how we perform the numerical quadrature of a matrix-valued integral. Since the bandstructure energies $\epsilon$ and $\bar\epsilon$ are both distributed with Gaussian weight functions, we employ Gaussian quadrature in each dimension to perform the integration.  We found, by benchmarking the equilibrium solution\cite{benchmark_equilib}, that averaging the results of two Gaussian quadratures with $N$ and $N+1$ points works better than choosing $2N+1$ points for the quadrature.  For most cases we discuss here, we use $N=54$ and $N+1=55$ points for the Gaussian quadrature routines, although some calculations were performed with the more accurate $N=100$ and $N+1=101$ routines. In the former case, we have a total of 5,941 quadrature points, while in the latter case it is 20,201 quadrature points.
Since the calculation of each matrix in the integrand of the
integral is independent of every other quadrature point, this part of the
code is easily parallelized.  The numerical quadrature requires two matrix multiplies and one matrix inversion for each quadrature point, and the integrand is a matrix. 

The calculation of the local Green's function from the local self-energy is the most computationally intensive part of the algorithm.  Fortunately that part parallelizes well in the master-slave format, since we need to send the slave nodes the self-energy matrix, and then send them the energies $\epsilon$ and $\bar\epsilon$ for the particular quadrature point, and the slaves calculate the integrand directly and accumulate the results locally. Then, once all bandenergies have been calculated, we use a recursive binary gather operation\cite{recursive} to accumulate the results and send them to the master efficiently.  This works by dividing the slave nodes in half and having one half send their accumulated results to the other half, and having those slave nodes accumulate the results they received with the ones they had.  A slave node that has sent its results to another slave node then becomes inactive. The procedure is repeated until only one slave node has all of the accumulated results, which is then sent to the master. The impurity solver which inputs the local Green's function for the lattice, and outputs the impurity self-energy,
is a serial code, that cannot be parallelized, because the matrix operations
need to all be performed in turn.  These calculations are performed on the master node and they require a number of matrix inversions and matrix multiplies, which are carried out with LAPACK and BLAS routines. We typically require between ten and one hundred iterations to reach convergence of the results (the total computer time for the
calculations presented here was about 600,000 cpu-hours on a Cray XT3,
900,000 cpu-hours on a SGI ALTIX, and 800,000 cpu-hours on a SUN OPTERON).  Overall, the code is quite efficient.  It has achieved over 65\% of the peak speed on 2032 cores of an SGI ALTIX machine, and scales with near linear scaling up to many hundreds to a few thousand cores (the main issue that affects the linear scaling is the fact that part of the algorithm is serial in nature and does not scale).

Once the Green's functions have converged, we calculate the current (in the Hamiltonian gauge) by evaluating the operator average
\begin{equation}
\langle {\bf j}(t)\rangle=-ei\sum_{\bf k}{\bf v}[{\bf k}+\theta(t)e{\bf E}t]G^<_{\bf k}(t,t).
\label{eq: current_def}
\end{equation}
The velocity component is $v_i({\bf k})=t^*\sin({\bf k}_i)/2\sqrt{d}$, and
all components of the current are equal when the field lies along the
diagonal (hence we can replace ${\bf v}_{\bf k}$ by $-\bar\epsilon$ and change the sum over momentum to a two-dimensional integral over $\epsilon$ and $\bar\epsilon$).
We also calculate the equal time retarded and lesser Green's functions
and their first two derivatives (at equal time) and compare those results to the exact
values~\cite{turkowski_freericks_2006}.  In general, these ``moments'' are
quite accurate as the step size is made smaller.

We find, nevertheless, that the results for the current and for the moments usually need to be extrapolated to the limit $\Delta t\rightarrow 0$. To do this, we
use a Lagrange interpolation formula to extrapolate the results to $\Delta t=0$.  Typically it is a quadratic extrapolation, requiring results at three different values of $\Delta t$, but sometimes we use higher order extrapolants.  This extrapolation procedure allows us to achieve quite accurate results for the moments, when compared to the exact values, and of the current (which must vanish when $t<0$). Further details of these
numerical issues and of the accuracies are presented 
elsewhere\cite{recursive,freericks_ugc_2007}.

We end our formalism discussion by defining the nonequilibrium many-body density of states.  First we convert from the time variables $t$ and $t^\prime$ to Wigner's average $T=(t+t^\prime)/2$ and relative $t_{rel}=t-t^\prime$ time variables.  Then the DOS is defined from the Fourier transform of the retarded Green's function with respect to the relative time.  In other words,
\begin{equation}
 G^R(T,\omega)=\int dt_{rel}e^{i\omega t_{rel}}G^R\left ( T+\frac{t_{rel}}{2},T-\frac{t_{rel}}{2}\right ) ,
\label{eq: gr_omega}
\end{equation}
is the retarded Green's function at each average time, and the many-body density of states satisfies
\begin{equation}
 \rho^{DOS}(T,\omega)=-\frac{1}{\pi}{\rm Im}G^R(T,\omega).
\label{eq: dos_def}
\end{equation}
We will use these relations to help analyze some of our results for the current.  Note that the nonequilibrium density of states can actually become negative for finite average times---it is nonnegative before the field is turned on (in equilibrium) and also in the limit $T\rightarrow\infty$ (the steady state).

\section{Numerical results}. 

We produce numerical calculations of the nonequilibrium
current as a function of time for the case of half-filling, where the conduction
electron and the localized electron fillings are each equal to 0.5.  In equilibrium, this system
has a metal-insulator transition at $U=\sqrt{2}$.
In the case where there is no scattering ($U=0$),
the Bloch oscillations continue forever.  In the presence of scattering (for metals), the
Bloch oscillations maintain the same approximate ``periodicity'', but the 
amplitude decays.  As the interactions increase further, the oscillations become irregular.

\begin{figure}[h]
\centering{
\includegraphics[width=3.3in,angle=0]{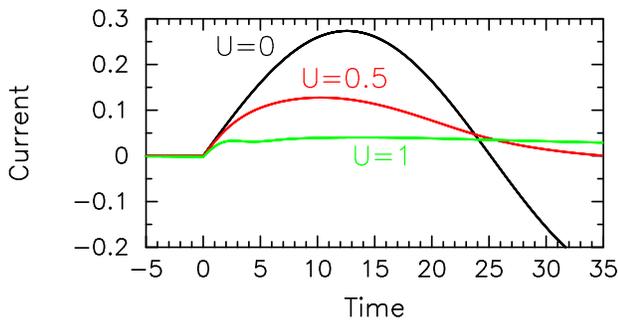}
}
\caption{
Scaled nonequilibrium current for different values of $U$ with $E=0.125$. All of these cases are metals in equilibrium. We used a quartic extrapolation formula with five $\Delta t$ values for $U=0.5$ ($\Delta t=0.1$, 0.067, 0.05, 0.04, and 0.033), while we used a quadratic extrapolation formula with three $\Delta t$ values for $U=1$ ($\Delta t=0.025$, 0.02, and 0.0167). (Color on-line.)
}
\label{fig: current_e=0.125}
\end{figure}

In Fig.~\ref{fig: current_e=0.125}, we plot the current for
the noninteracting case, the case of a strongly scattering metal $(U=0.5$, red),
and the case of an anomalous metal $(U=1$, green; anomalous in the sense that there is a dip in the many-body density of states near the chemical potential) for $E=0.125$.  The Bloch period in this case is $16\pi\approx 50$. The time range we can extend the calculations out to is $t_{max}=35$, so we do not see one full Bloch oscillation in the time window. The initial temperature of the system satisfied $\beta=10$, and the field is turned on at $t=0$.  We checked the errors of the extrapolated data against the known moment sum rules.  The errors are largest at small times (in equilibrium, before the field is turned on) and improve for larger times.  As a benchmark, we record the maximal error in the moments for times greater than $t=5$.  The exact moments are equal to 1 for the zeroth moment and $0.5+U^2/4$ for the second moment. The $U=0.5$ case has errors in the zeroth moment less than 1\% and errors of 2\% for the second moment.  The $U=1$ case has errors less than 1\% for the zeroth moment and errors of 3\% for the second moment.  We find generically, that the calculations require smaller $\Delta t$ values for smaller electric fields.

In the strongly scattering metal, the Bloch oscillations appear to be simply damped, while in the anomalous metal, we start to see additional oscillations develop at short time, which appear to disappear at longer times. While we have clearly not reached the steady state yet, it does appear that the current is approaching a constant nonzero value for the anomalous metal at long times, as we expect for a driven correlated material.

\begin{figure}[h]
\centering{
\includegraphics[width=3.3in,angle=0]{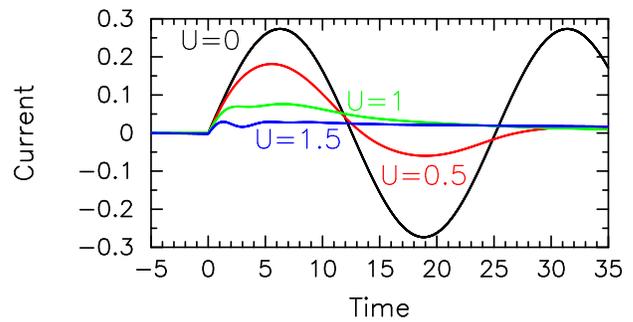}
}
\caption{
Scaled nonequilibrium current for different values of $U$ (0, 0.5, 1, and 1.5) with $E=0.25$. All of these cases are metals in equilibrium except for $U=1.5$ which is a near-critical Mott insulator. We used a quadratic extrapolation formula with three $\Delta t$ values for $U=0.5$ ($\Delta t=0.1$, 0.067 and  0.05, 0.04, and 0.033), $U=1$ ($\Delta t=0.05$, 0.04, and 0.033),
and $U=1.5$ ($\Delta t=0.033$, 0.025, and 0.02). (Color on-line.)}
\label{fig: current_e=0.25}
\end{figure}

In Fig.~\ref{fig: current_e=0.25}, we plot the scaled current for $E=0.25$ and four $U$ values ranging from a ballistic metal to a near critical Mott insulator.  The Bloch period is now approximately 25 units, so we fit about one and one half oscillations in our time window. The behavior for these cases is similar in many respects to what we saw for the smaller field.  The oscillations are damped and as the scattering strength increases we see some additional shorter period wiggles develop, especially for short times. The larger $U$ values appear like they are approaching a steady-state limit (with a smaller value of the current than for $E=0.125$, as expected for large fields\cite{secondorderpert}), but it is also clear that the larger field strength increases the transient response and makes it more difficult to reach the steady state within our time window.  We also find that our accuracy improves and we do not need to employ as small values of $\Delta t$; this is a general trend that continues as we increase the electric field.  The scaled moments show the following accuracies: $U=0.5$, less than 1\% error for the zeroth and second moments; $U=1$, less than 1\% error for the zeroth and second moments; and $U=1.5$, less than 2\% error for the zeroth moment and less than 4\% error for the second moment.

\begin{figure}[h]
\centering{
\includegraphics[width=3.3in,angle=0]{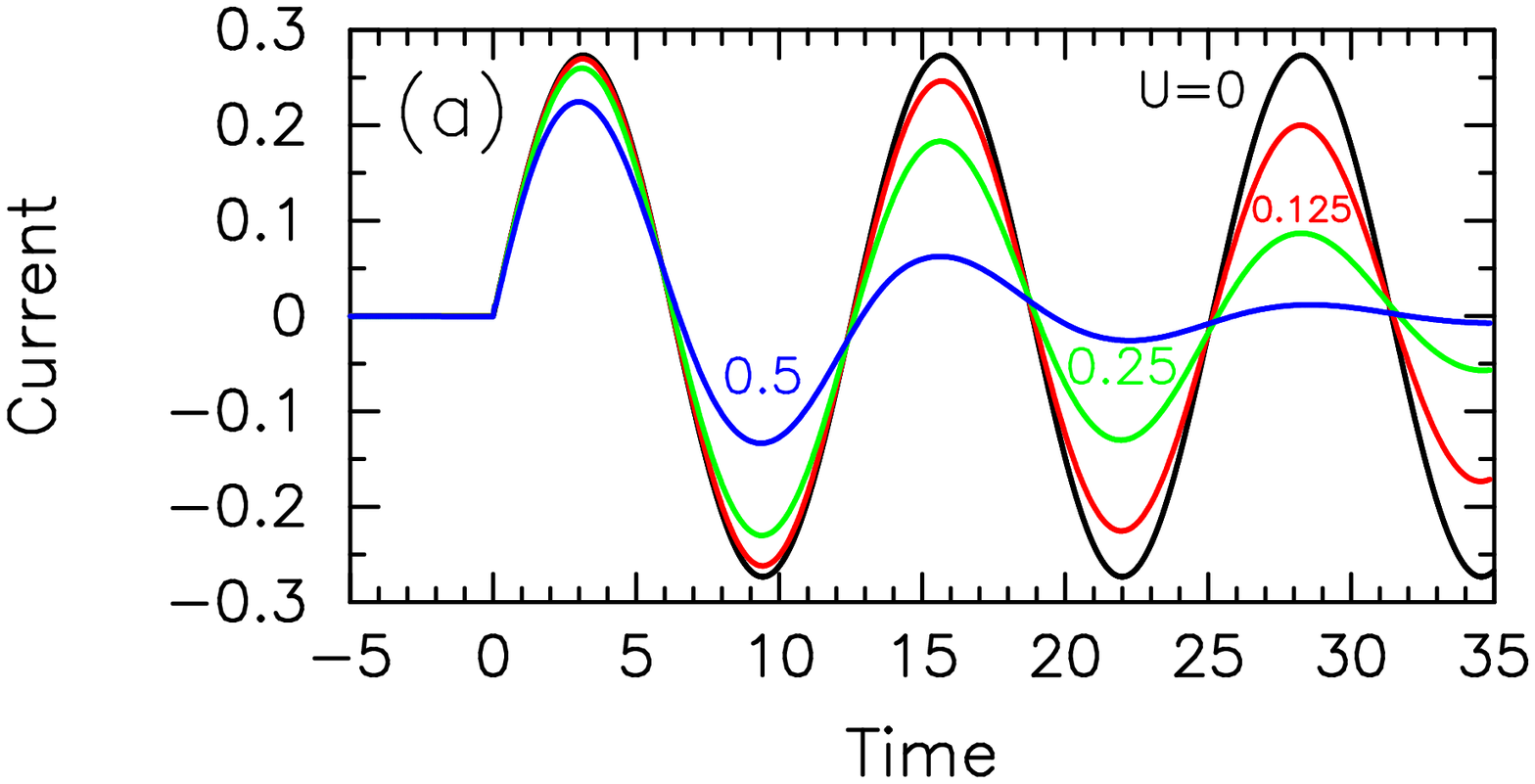}
}
\centering{
\includegraphics[width=3.3in,angle=0]{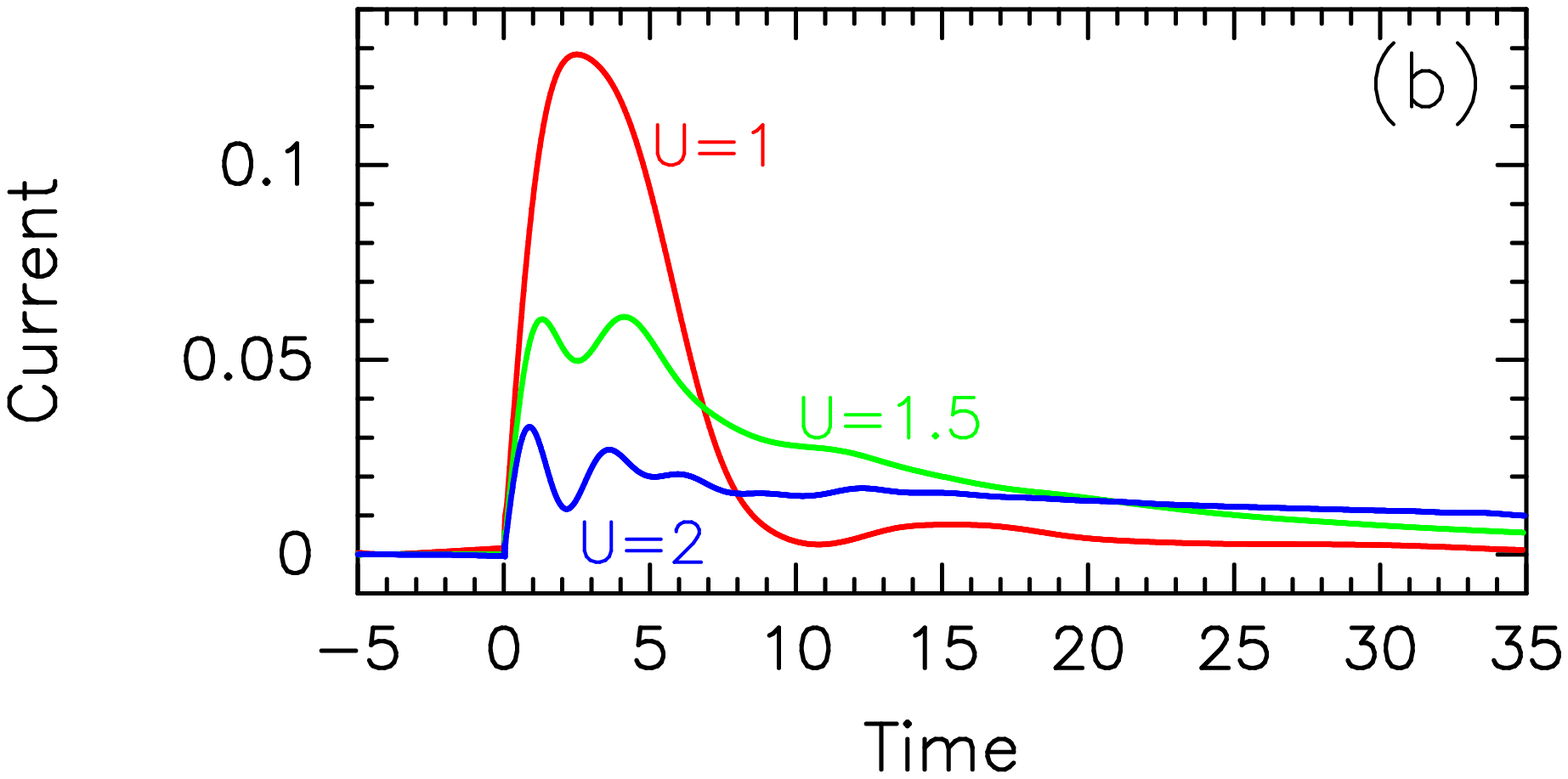}
}
\caption{
Scaled nonequilibrium current for different values of $U$ with $E=0.5$. In panel (a), we show cases that are metals in equilibrium. In panel (b), we show one anomalous metal ($U=1$) and two Mott insulators ($U=1.5$ and $U=2$); note the reduced scale for the vertical axis versus that in panel (a). We used a quadratic extrapolation formula  for $U\le 1$ ($\Delta t=0.1$, 0.067, and 0.05 for $U=0.125$, $U=0.25$ and $U=0.5$; $\Delta t=0.067$, 0.05, and 0.04 for $U=1$), a cubic extrapolation formula for $U=1.5$ ($\Delta t=0.067$, 0.05, 0.04, and 0.033), and a quartic extrapolation formula for $U=2$ ($\Delta t=0.05$, 0.04, 0.033, 0.025, and 0.02). (Color on-line.)
}
\label{fig: current_e=0.5}
\end{figure}

In Fig.~\ref{fig: current_e=0.5}, we plot the $E=0.5$ current for normal metals in panel (a) and for the anomalous metal and two Mott insulators in panel (b).  Here we have nearly three full Bloch oscillations within our time window, and the metallic phases are behaving pretty much as we would expect---they show Bloch oscillations with the same period and are increasingly damped as $U$ increases.  Unfortunately, the larger field drives the system for a longer period of time, so we are unable to see the steady state limit emerging for any of these metallic cases.  While we expect the steady state current to be a constant {\it dc} response, we cannot rule out the possibility of a small amplitude oscillatory response as well. The anomalous metal and insulating cases in panel (b) also display interesting behavior.  For example, we see one prominent oscillation for the first half Bloch period and then the response is sharply damped afterwards, but the oscillations have no regularity to them, and as the system becomes more insulating we see additional shorter period oscillations develop, similar to what occurred for smaller fields, but here they are more prominent.  For longer times, the response is not yet reaching a steady state, although it is flattening out much faster.  Notice that in some cases, the long-time current response is {\it larger} for more strongly interacting systems (notice how the $U=2$ response lies above the $U=1.5$ and $U=1$ responses for long times).  We will see this phenomenon recur as we increase the electric field further.
As mentioned above, the accuracy continues to improve as we increase the field. For all cases shown, the error in the zeroth and second moments is less than 1\% for times larger than 5.

\begin{figure}[h]
\centering{
\includegraphics[width=3.3in,angle=0]{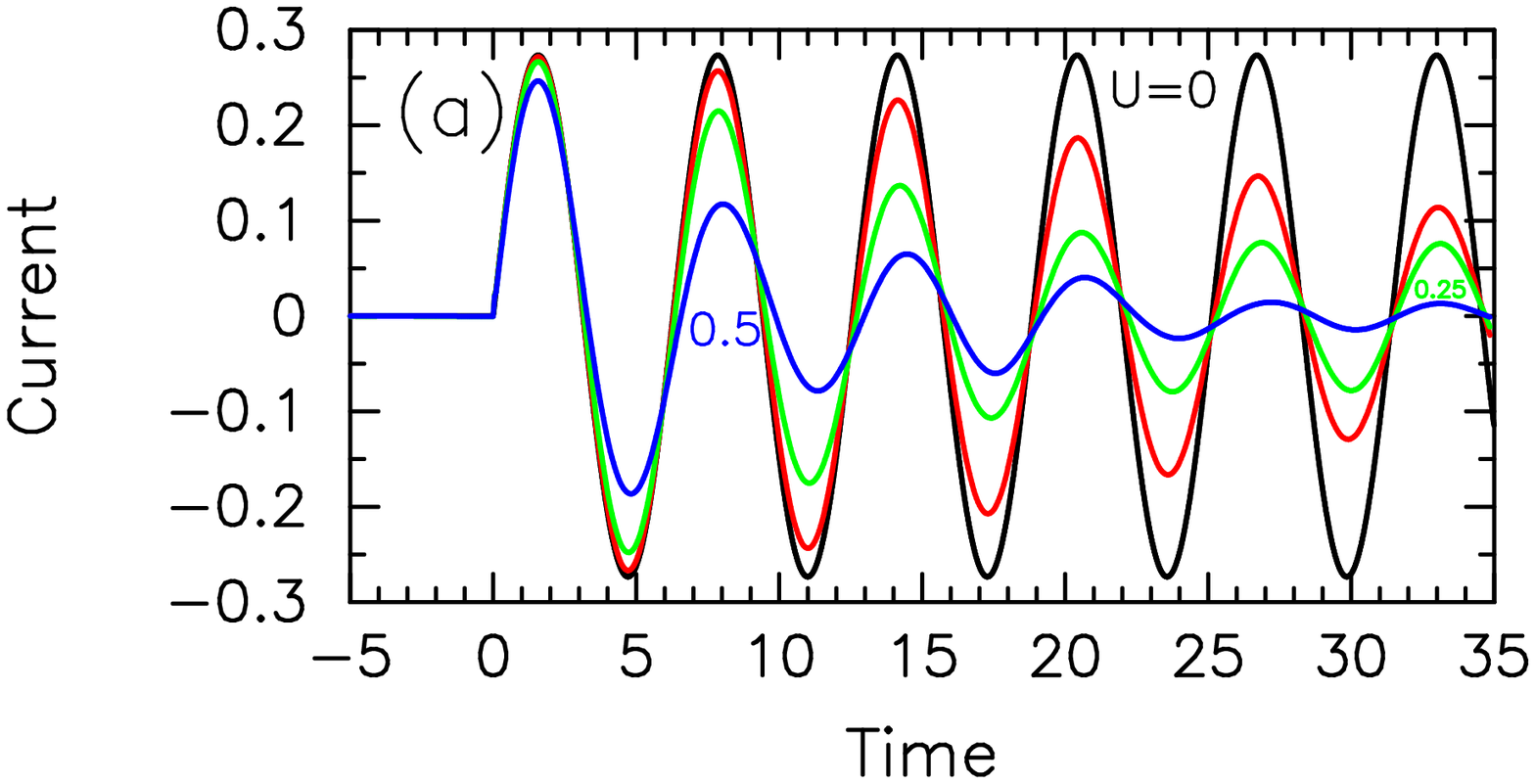}
}
\centering{
\includegraphics[width=3.3in,angle=0]{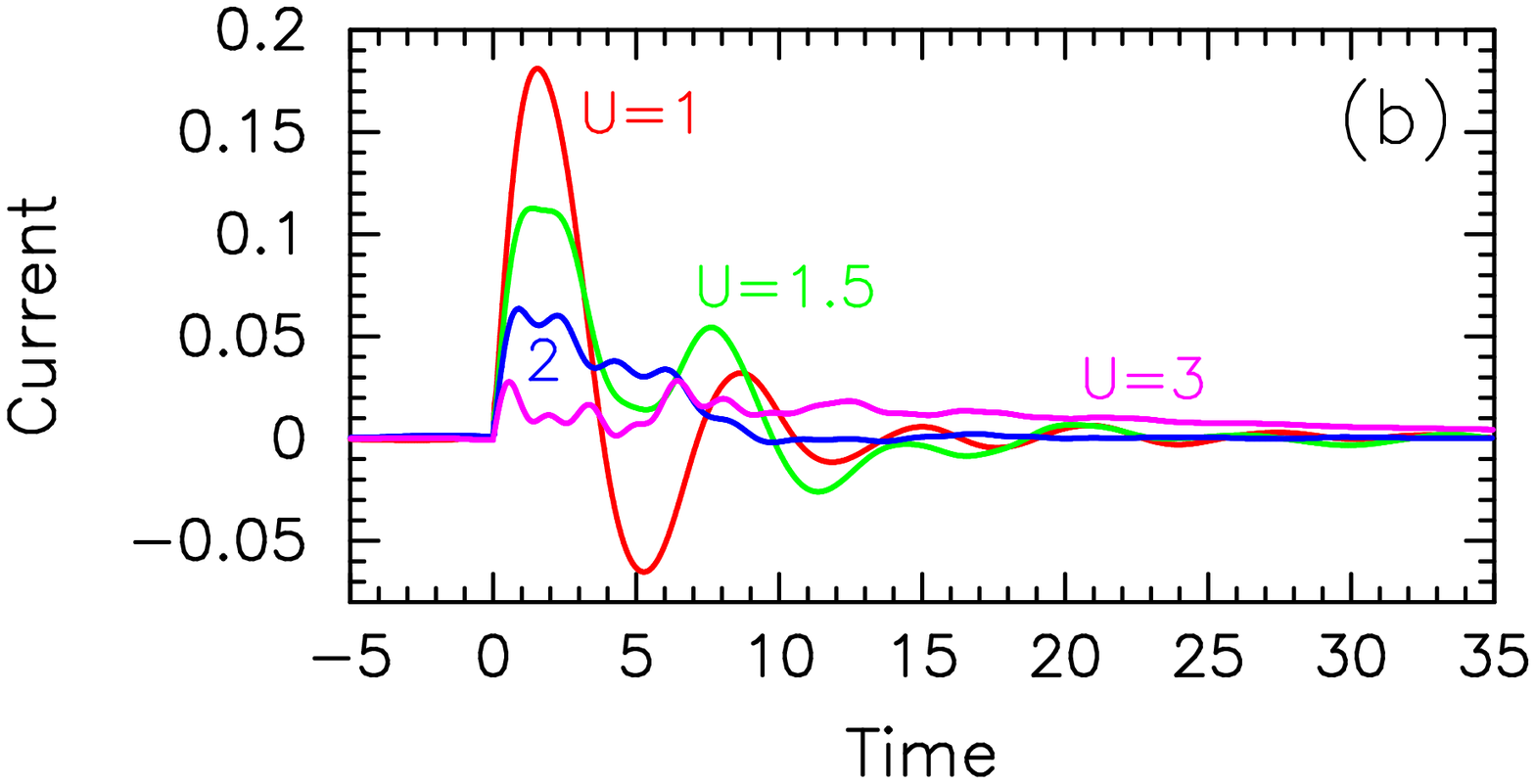}
}
\caption{
Scaled nonequilibrium current for different values of $U$ with $E=1$. In panel (a), we show cases that are metals in equilibrium ($U=0$, 0.125, 0.25, and 0.5. In panel (b), we show one anomalous metal ($U=1$) and three Mott insulators ($U=1.5$, 2 and 3); note the reduced scale for the vertical axis versus that in panel (a). We used a quadratic extrapolation formula  for $U\le 0.5$ ($\Delta t=0.1$, 0.067, and 0.05), a cubic extrapolation formula for $U=1$ ($\Delta t=0.1$, 0.067, 0.05, and 0.04), a quadratic extrapolation for $U=1.5$ ($\Delta t=0.067$, 0.05, and 0.04), a quartic extrapolation formula for $U=2$ ($\Delta t=0.067$, 0.05, 0.04, 0.033, and 0.025), and a quadratic extrapolation formula for $U=3$ ($\Delta t=0.02$, 0.0167, and 0.0143). (Color on-line.)
}
\label{fig: current_e=1}
\end{figure}

In Fig.~\ref{fig: current_e=1}, we plot results for the current with $E=1$.  In panel (a) we show normal metals $U=0$, 0.125, 0.25, and 0.5, while in panel (b) we show one anomalous metal $U=1$ and three Mott insulators ($U=1.5$, 2, and 3).  Note how the metals continue to show damped Bloch oscillations, and that we cannot reach the steady state within the time window. It may appear that the $U=0.5$ case is not oscillating with the Bloch period, but that deviation is most likely coming from some higher harmonics in the transient response that should vanish as time increases, because the oscillations are clearly not with some slightly larger period in this range. The accuracy for all of these metallic cases is better than 1\% for both the zeroth and second moment for times greater than $t=5$. 

Panel (b) shows the current for Mott insulators (and one anomalous metal). As the system becomes more Mott-insulating, we see more irregular oscillations of a larger amplitude.  Note how, in addition, the $U=3$ Mott-insulating current lies above the $U=2$ Mott-insulating current at long times.  For smaller $U$ values, the oscillations still maintain a fair amount of regularity with a period close to the Bloch period.  As shown in Ref.~\onlinecite{freericks_prl_2006}, the irregular oscillations continue out to long times.  These results clearly show the emerging trends as we move from a metal to a Mott insulator.  The current initially is quenched via a damping of the Bloch oscillations, but as the coupling strength increases, the oscillations become more irregular and lower in overall amplitude.  As the field strength increases, we see the oscillations survive out to longer and longer times.  In most cases, we are unable to reach the steady-state limit within our time window. The accuracy of our calculations is still quite good.  The extrapolated results have errors of less than 1\% for for both moments when $U=1$, less than 2\% for both moments when $U=1.5$ and $U=2$, and less than 3\% for both moments when $U=3$.

\begin{figure}[h]
\centering{
\includegraphics[width=3.3in,angle=0]{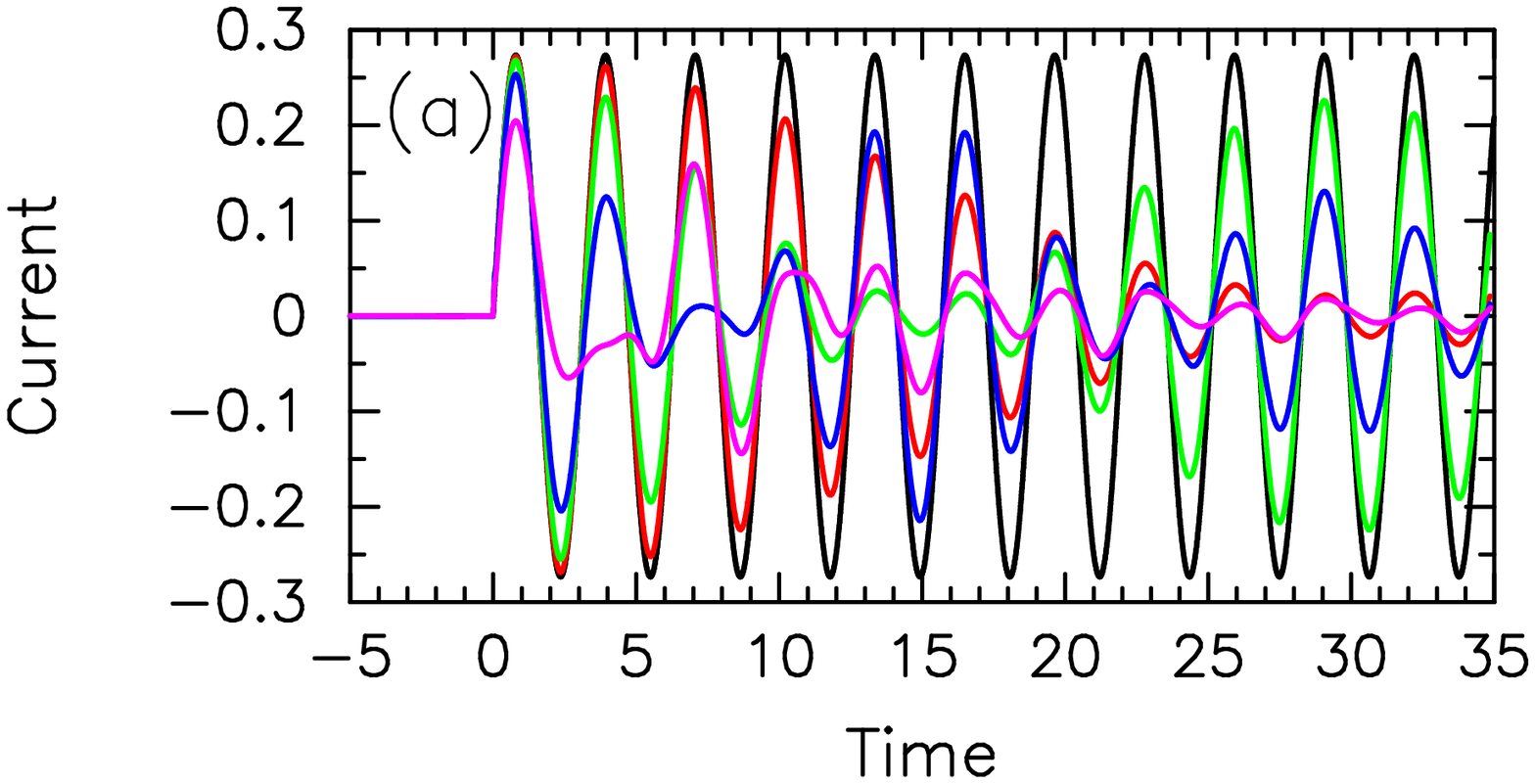}
}
\centering{
\includegraphics[width=3.3in,angle=0]{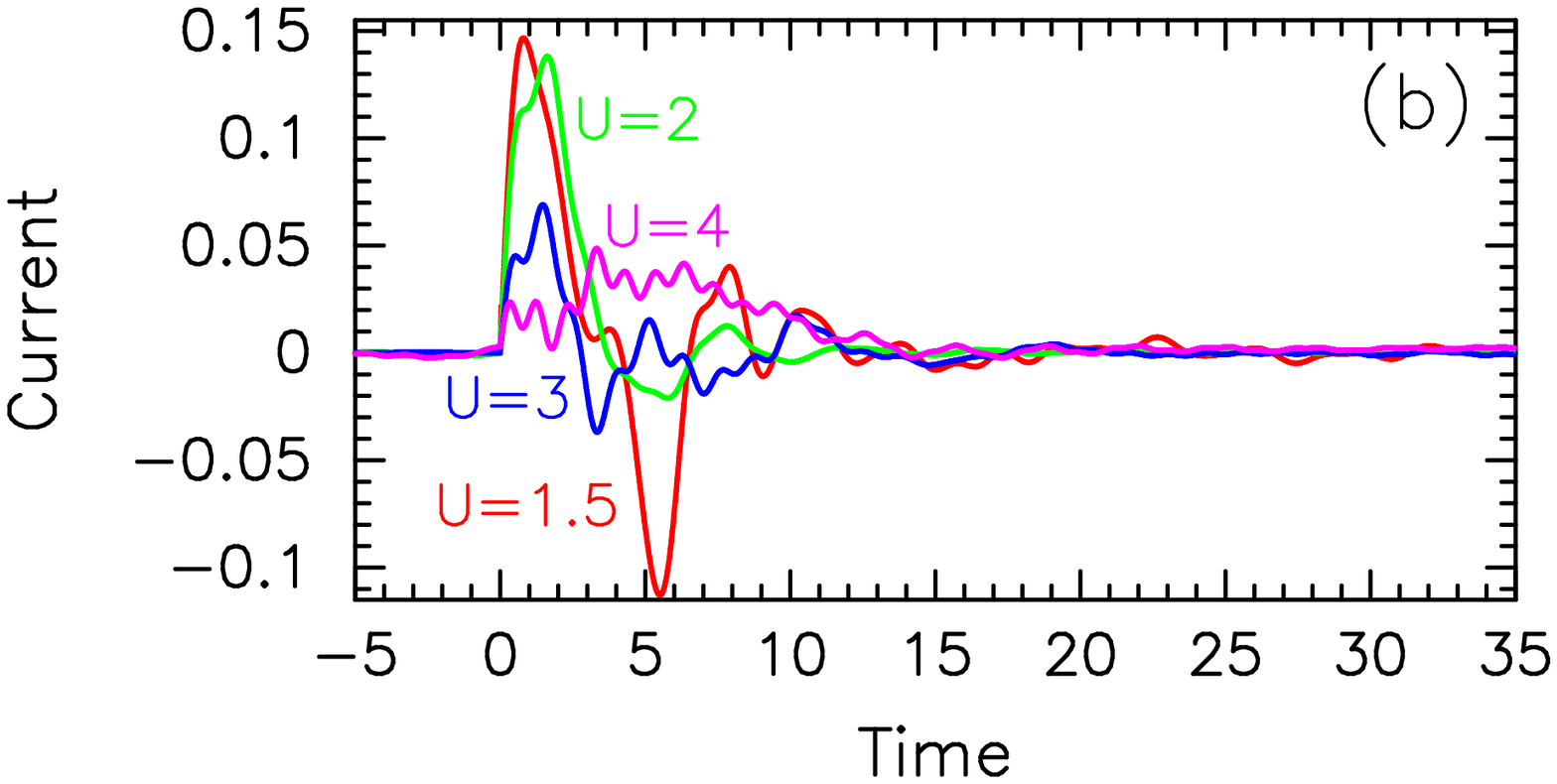}
}
\caption{
Scaled nonequilibrium current for different values of $U$ with $E=2$. In panel (a), we show cases that are metals in equilibrium ($U=0$, 0.125, 0.25, 0.5 and 1). In panel (b), we show  Mott insulators ($U=1.5$, 2, 3, and 4); note the reduced scale for the vertical axis versus that in panel (a). We could not fit the $U$ labels onto panel (a). The $U=0$ case (black) has no damping to the oscillations.  The $U=0.125$ case (red) looks like a damped oscillator, because the beat period is about 50.  The $U=0.25$ case (green) shows one beat period (it has the second largest amplitude near $t=30$).  The $U=0.5$ case (blue) shows a few beats and is the third largest amplitude near $t=30$.  The $U=1$ case (magenta) is the most irregular looking of the curves. We used a quadratic extrapolation formula  $U\le 0.5$ ($\Delta t=0.1$, 0.067, and 0.05), a cubic extrapolation formula for $U=1$ ($\Delta t=0.1$, 0.067, 0.05, and 0.04), and for $U=1.5$ ($\Delta t=0.067$, 0.05, 0.04, and 0.033), a quadratic extrapolation formula for $U=2$ ($\Delta t=0.04$, 0.033, and 0.025), a cubic extrapolation formula for $U=3$ ($\Delta t=0.04$, 0.033, 0.025, and 0.0167), and a linear extrapolation for $U=4$ ($\Delta t=0.0167$ and 0.0143). (Color on-line.) 
}
\label{fig: current_e=2}
\end{figure}

In Fig.~\ref{fig: current_e=2}, we plot the current for our final electric field $E=2$. The behavior in this case is quite different from the other cases. In the metallic case, we do not have simple damped oscillations.  Instead, the oscillations initially decay but then grow, reminiscent of damped beats.  The beat period turns out to be $2\pi/U$, decreasing as $U$ increases.  This result was predicted from nonequilibrium calculations for two particles on a one-dimensional chain\cite{beats}, but we see here that it appears to only enter once the electric field is large enough (for an infinite-dimensional lattice).  We will treat this phenomenon in more detail below and resolve the underlying physical mechanism behind its occurrence. The beat period becomes so short, that the $U=1$ results look like quite irregular oscillations.

\begin{figure}[h]
\centering{
\includegraphics[width=3.3in,angle=0]{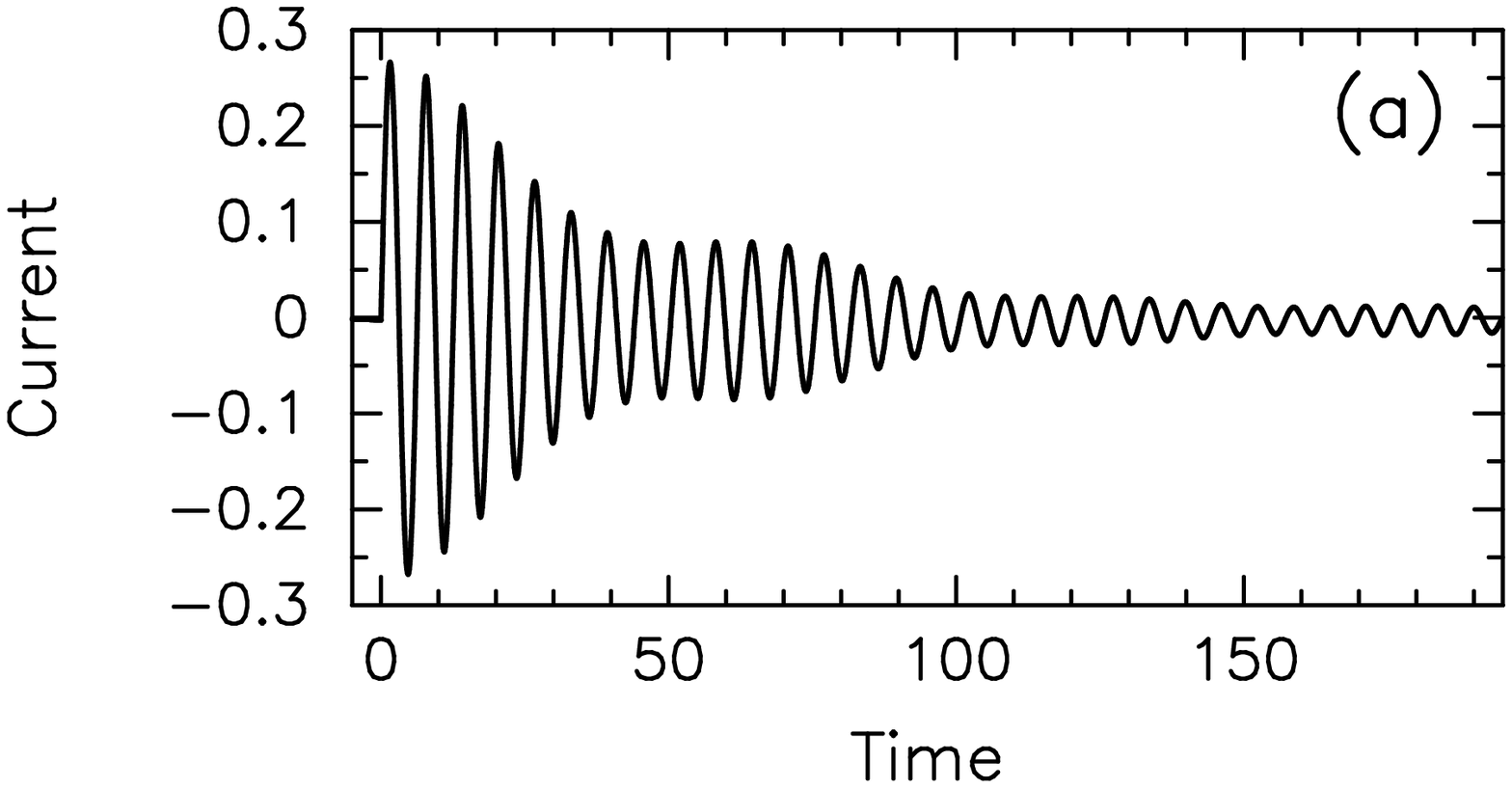}
}
\centering{
\includegraphics[width=3.3in,angle=0]{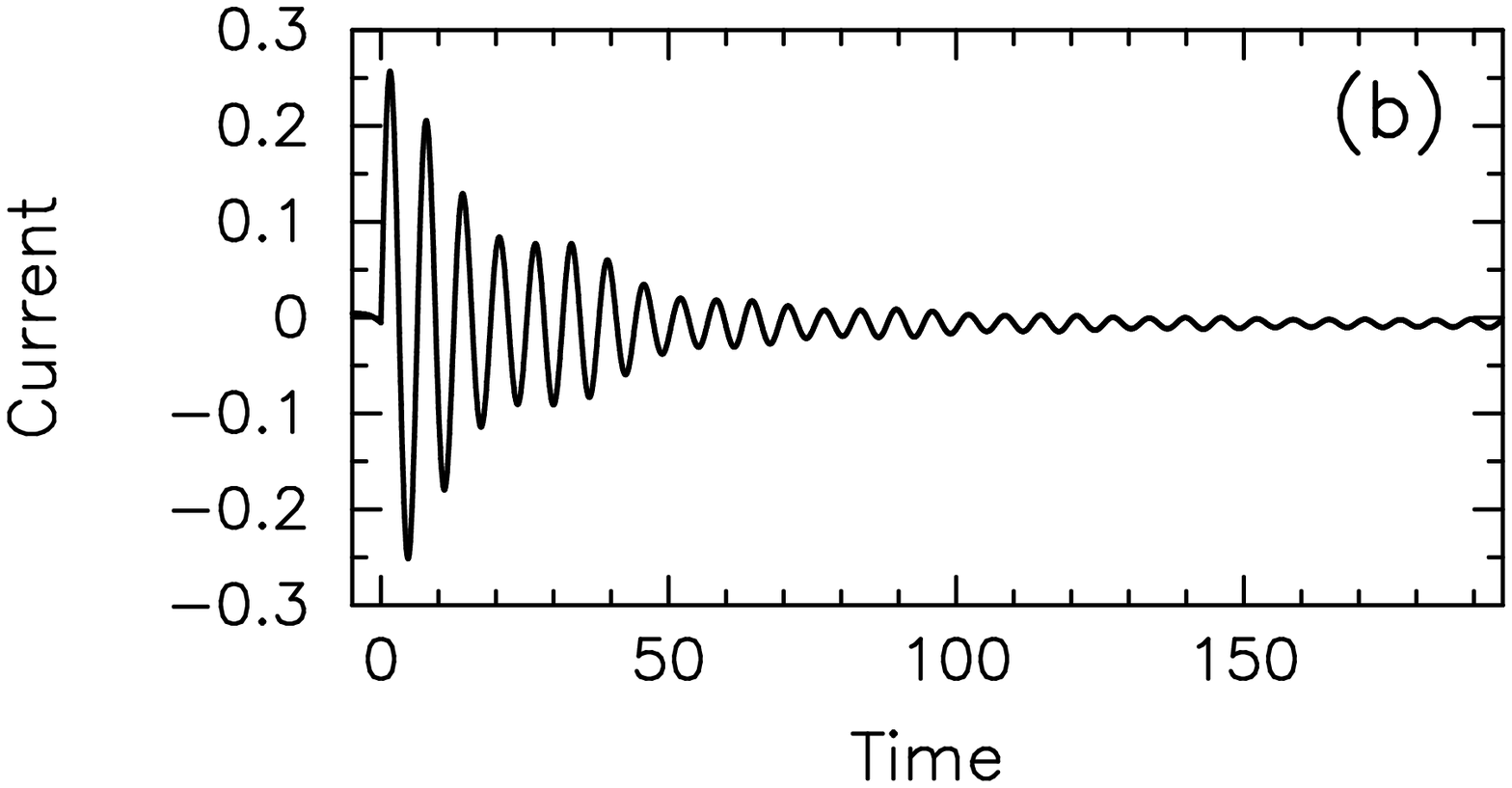}
}
\centering{
\includegraphics[width=3.3in,angle=0]{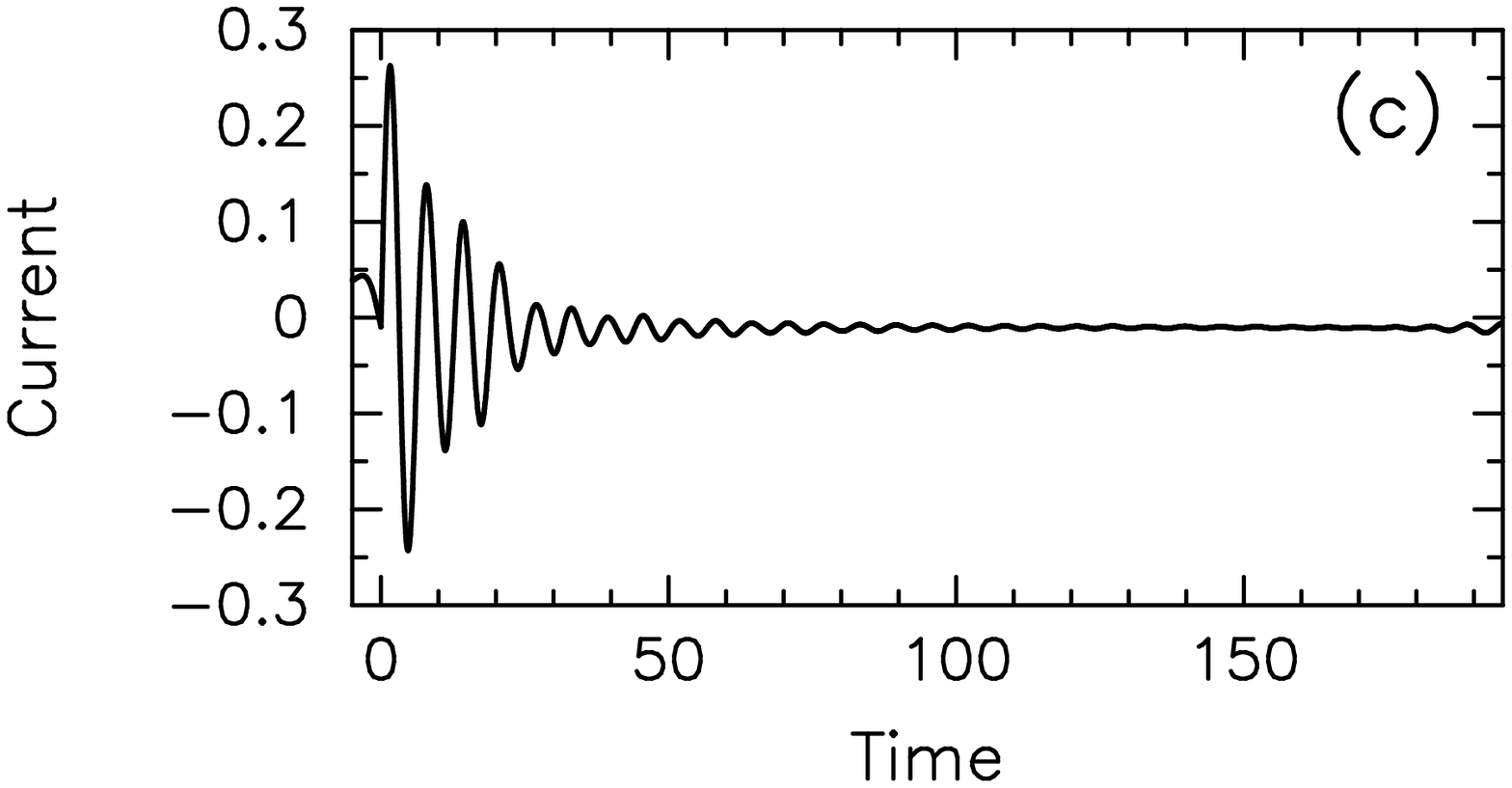}
}
\caption{
Unscaled nonequilibrium current ($\Delta t=0.1$) for $E=1$ and (a) $U=0.125$, (b) $U=0.25$, and (c) $U=0.5$.  The longer time cutoff allows us to see the initial development of beat-like behavior in the current response.
}
\label{fig: beats_e=1}
\end{figure}

The beats disappear as we pass through the metal-insulator transition.  Instead, we see the irregularity in the oscillations become more apparent.  In panel (b), one can see that the near-critical Mott insulator ($U=1.5$) still looks like it could be respresented in terms of damped oscillations with beats, but the oscillations are not simple sinusoidal oscillations. As $U$ is increased further, the oscillations are sharply attenuated for times larger than 15. They continue to have significant short-period irregular oscillations, which appear to develop more strongly as $U$ increases. In addition, we see all of the different Mott insulators to have small current at long times, and we do not see some cases having significantly larger current as we did when $E=1$. In spite of our ability to extrapolate our numerical calculations to $\Delta t\rightarrow 0$, these Mott-insulating results are the most difficult to obtain reliable data for, and it is quite likely that the final curves for the current have pointwise errors on the order of 10\%, but we have no way of rigorously estimating the errors, because our calculations have been pushed to the limit of what we can reliable achieve.  The errors for the metallic cases are all quite low (less than 1\% for both moments for all metals and for $U=1.5$).  But they grow in the Mott insulators (less than 5\% error for the first moment, and less than 7\% error for the second moment for larger $U$ values).

\begin{figure}[h]
\centering{
\includegraphics[width=3.3in,angle=0]{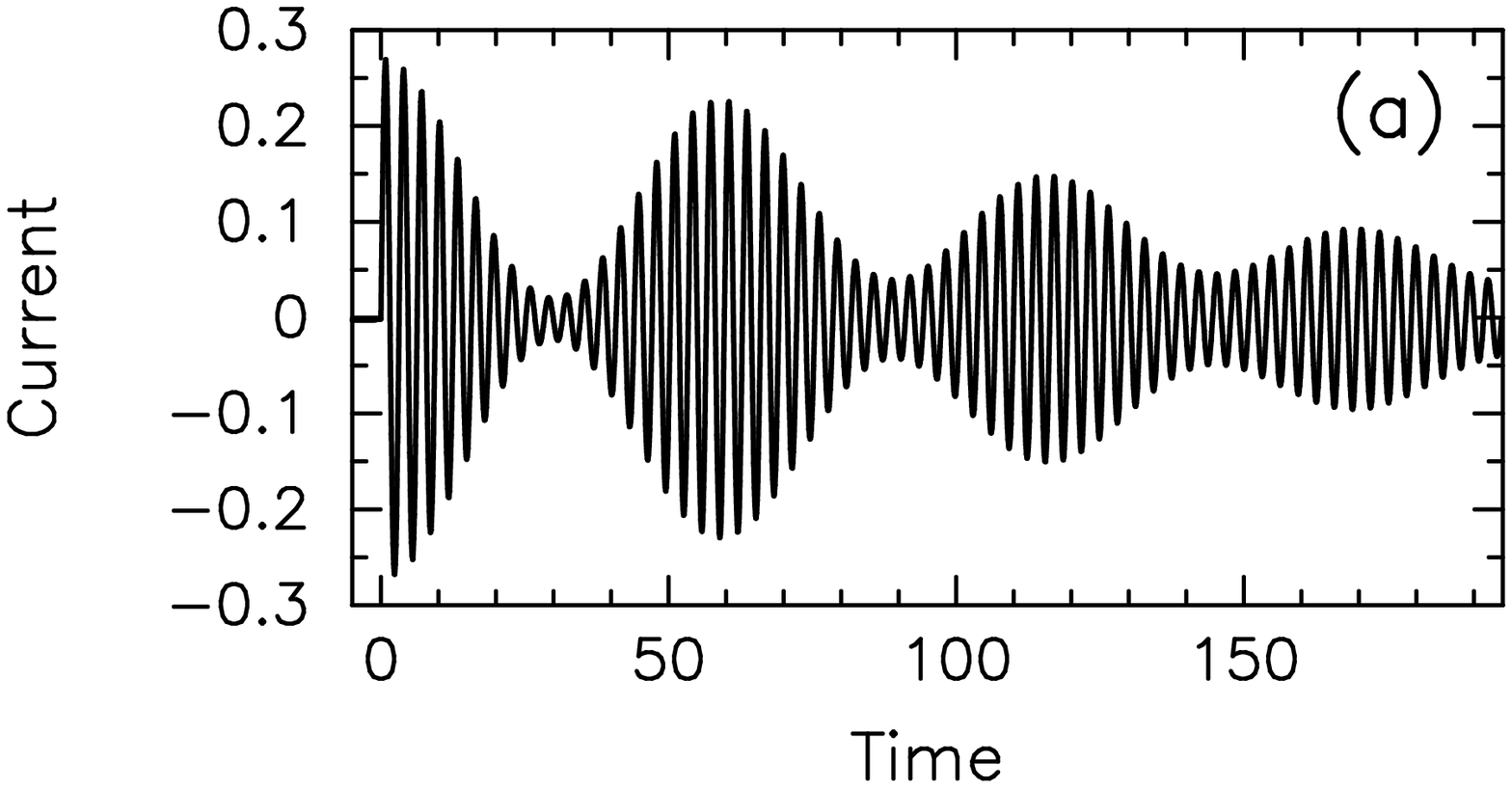}
}
\centering{
\includegraphics[width=3.3in,angle=0]{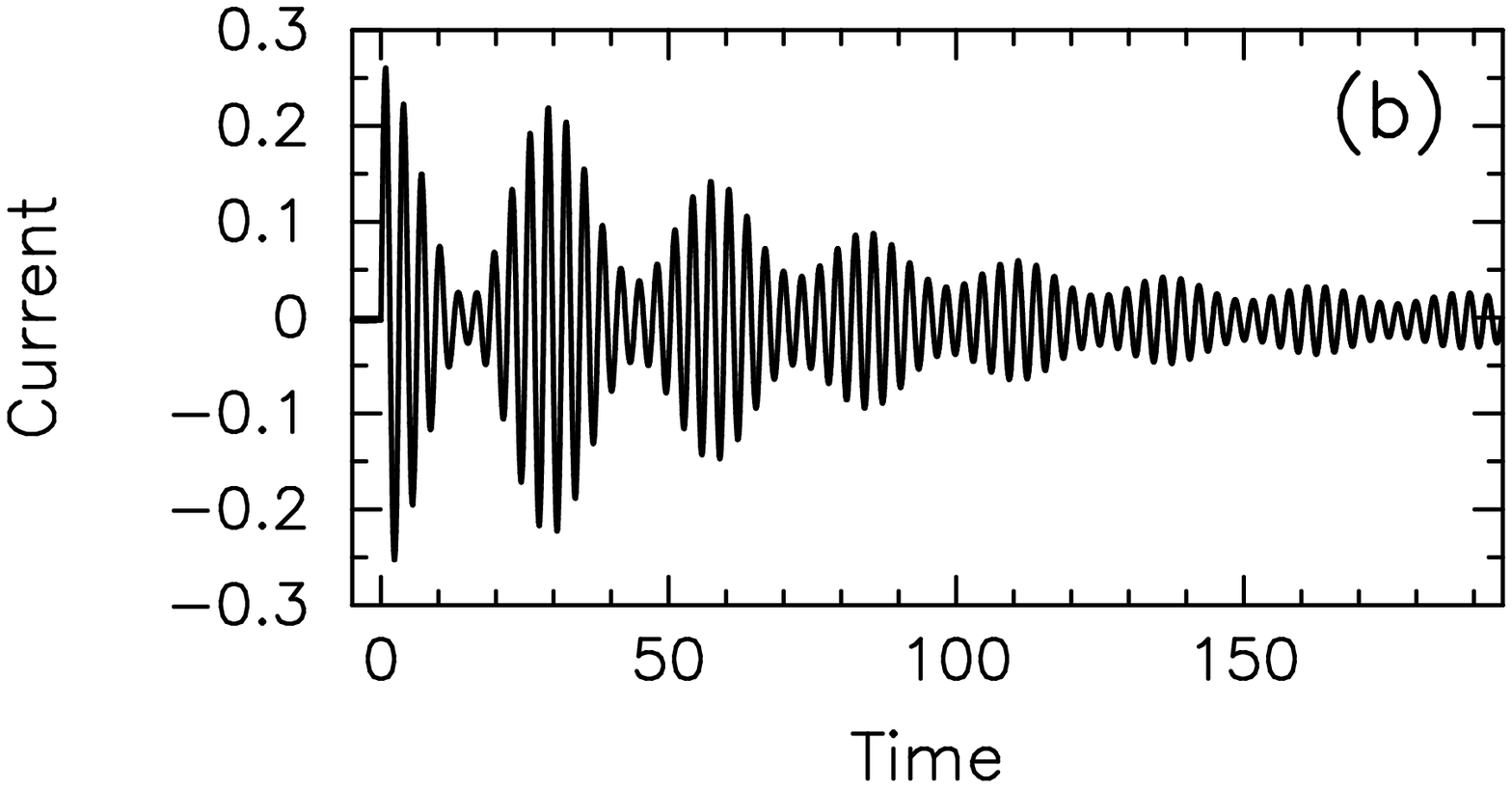}
}
\centering{
\includegraphics[width=3.3in,angle=0]{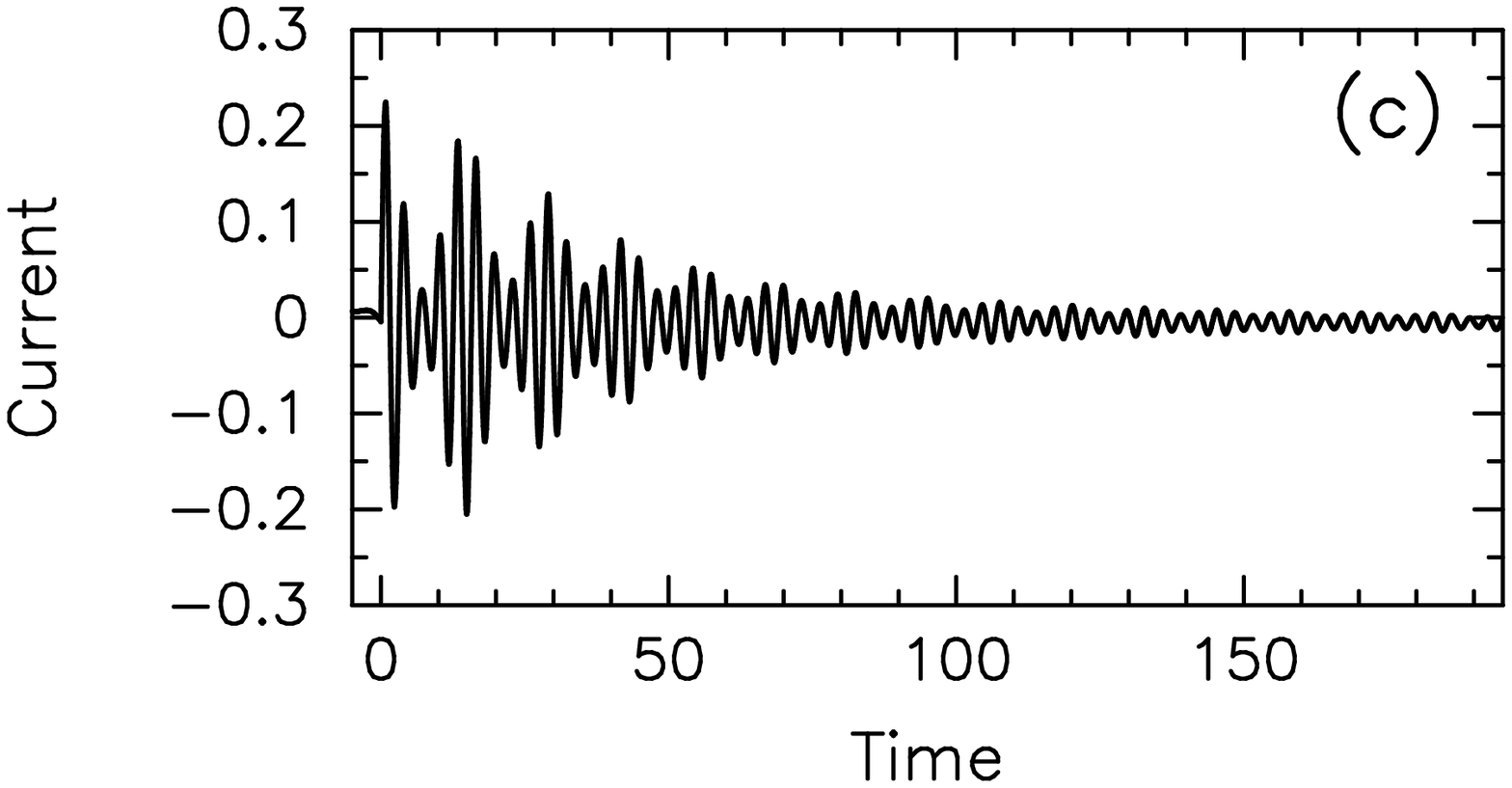}
}
\caption{
Unscaled nonequilibrium current ($\Delta t=0.1$) for $E=2$ and (a) $U=0.125$, (b) $U=0.25$, and (c) $U=0.5$.  Here the beats are clearly visible, but one can also see a dephasing at long times, because the beat amplitude does not go to zero.
}
\label{fig: beats_e=2}
\end{figure}

Because the phenomena of beats newly develops for large electric fields in metals, we need to investigate the origin of the beats more thoroughly.  To do this, we perform a series of calculations with $\Delta t=0.1$ and no scaling of the data, but with a much longer time window ($t$ runs out to 195). We ran calculations for $E=0.5$, $E=1$, and $E=2$.  In the case when $E=0.5$, we saw no indication of any beat phenomena in the data.  The results for $E=1$ are shown in Fig.~\ref{fig: beats_e=1}.  Panel (a) is for $U=0.125$, panel (b) for $U=0.25$, and panel (c) for $U=0.5$.  Note how there is clearly some behavior that is reminiscent of beats in this data (especially for small $U$), but it is not well formed beats yet. [Note that the increase in amplitude of the current at long times in panel (c) is a systematic error associated with the large discretization size for the $U$ value and the boundary in time.] We do see the period of the beat-like phenomena to decrease as $U$ increases, and it appears to be equal to $2\pi/U$.  When we move to the larger field of $U=2$, which is plotted in Fig.~\ref{fig: beats_e=2}, the beats become completely transparent, even if the amplitude modulation becomes somewhat dephased at long times.  We clearly see the beat period decreasing as $U$ increases and being inversely proportional to $U$,

\begin{figure}[h]
\centering{
\includegraphics[width=3.3in,angle=0]{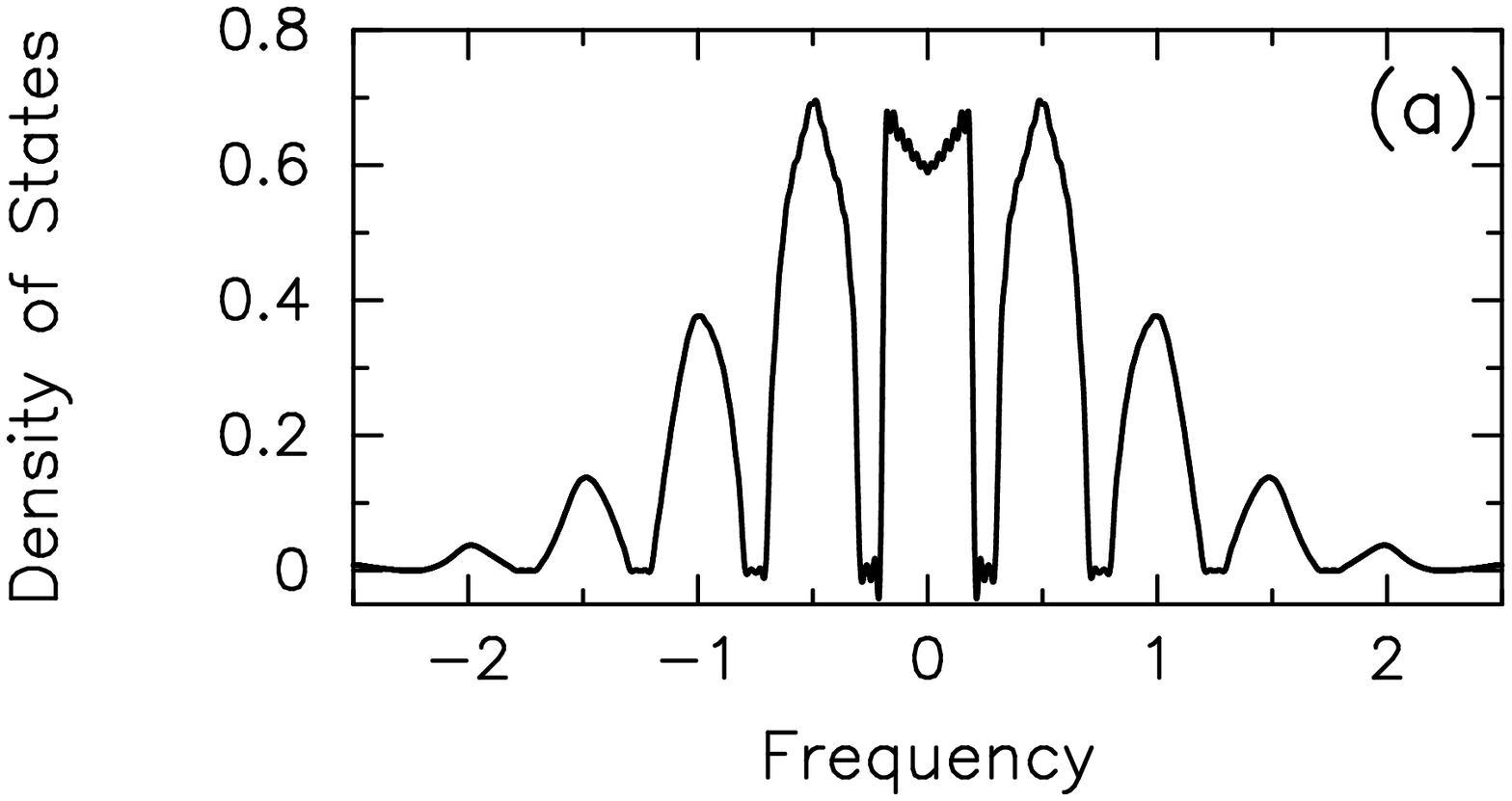}
}
\centering{
\includegraphics[width=3.3in,angle=0]{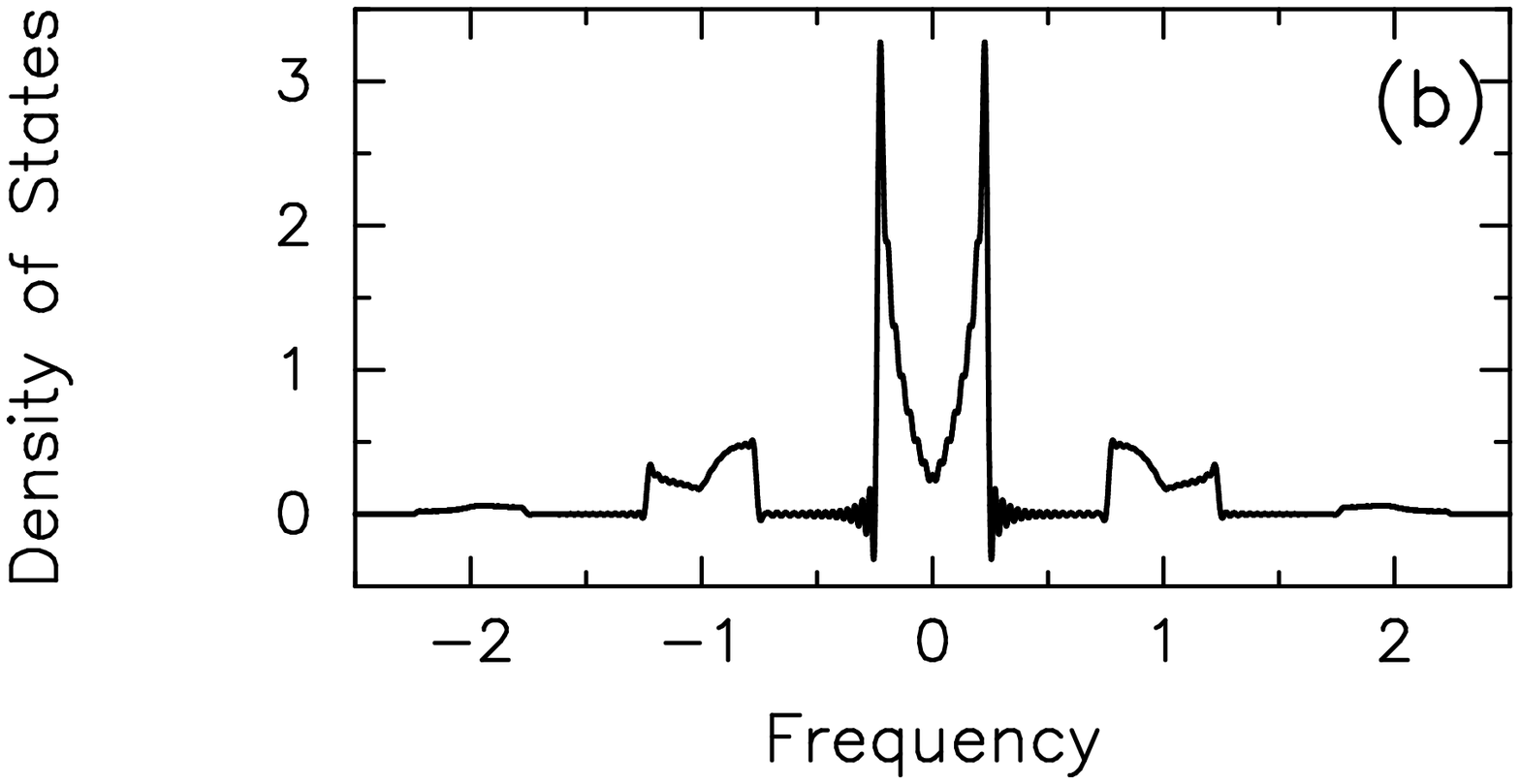}
}
\centering{
\includegraphics[width=3.3in,angle=0]{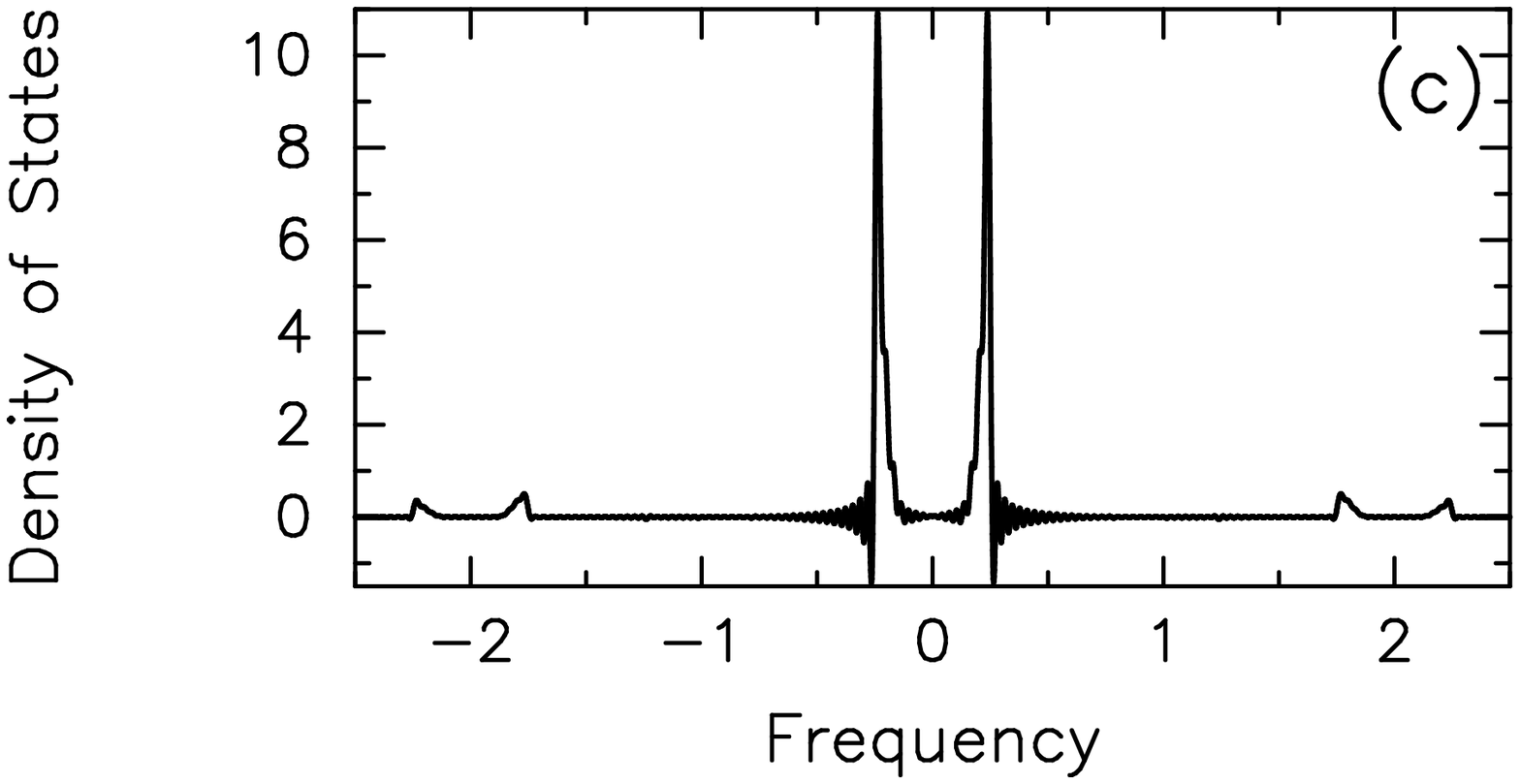}
}
\caption{
Local many-body density of states with $U=0.5$ for time $T=95$ units after the electric field was turned on and (a) $E=0.5$, (b) $E=1$, and (c) $E=2$; the data are from unscaled results with $\Delta t=0.1$. The Wannier-Strak ladder would have delta functions at the Bloch frequencies $nE$. Here the many-body density of states is broadened into bands about the Bloch frequencies with a width of approximately $U$.  Note how the central band becomes sharply peaked at $\omega=\pm U/2$ as $E$ increases.  This is the source of the two frequencies separated by $U$ that creates the beats in the current as a function of time.
}
\label{fig: dos}
\end{figure}

One can ask what the cause of these beats can be?  Since the beat period is $2\pi/U$, we immediately suspect the result is coming from two frequencies equal to the Bloch frequency plus or minus $U/2$. Then the combination of these oscillations will produce the expected beats.  If we extract the retarded Green's function from the contour-ordered Green's function, transform to Wigner coordinates and perform a Fourier transform with respect to the relative time, we can then plot the local many-body density of states as a function of average time.  For large times, the density of states approaches a steady state limit, and we plot a large time result ($T=95$) in Fig.~\ref{fig: dos} for $U=0.5$ and the three different $E$ values (0.5, 1, and 2).  One can see that as the electric field is increased, the DOS develops two sharp peaks centered at $\omega=\pm U/2$.  While we can see these peaks already for $E=1$, they really become well separated and distinct for $E=2$.  We believe that these two delta-function-like peaks, when combined with the underlying Bloch frequency, are the source of the beat phenomena in the current, and since we do not see the double peak structure develop for smaller fields, the beat behavior in the current is a strong field behavior.  This is hinted at in the work with two particles in one-dimension\cite{beats}, but that work showed the additional oscillations occurring for all $U$ and $E$.  As one increases the particle density, and the spatial dimension, the occurence of beats requires a critical value of the electric field, which is around $E=1$ for our case.

\section{Conclusions}

In this work, we have shown a detailed derivation of the generalization of DMFT to nonequilibrium problems given by the so-called Keldysh boundary condition---the system starts in equilibrium and then a field is turned on and the system is monitored as it is driven to a steady state. Our focus was on the current and how the Bloch oscillations are quenched and change character as electron-electron correlations are increased.  We find for weak fields the picture in metals is pretty much what one might have guessed---the Bloch oscillations are damped and the system approaches a steady state with what appears to be a constant current (although we cannot rule out low amplitude oscillations with the Bloch period superimposed on top of the constant response). As the field is increased, we find the oscillations survive out to longer and longer times.  When the coupling increases to that of a Mott insulator, the oscillations change their character and become quite irregular.  This process is a slow evolution and not a sharp ``phase transition''. Most interesting, is the appearance of beats in the current for large electric fields in metals.  These beats have a period inversely proportional to the interaction strength and are quite robust. As time increases however, the beats do dephase, and eventually we get a constant amplitude response (which may be decreasing due to damping) that no longer has an oscillatory envelope.  
We also explained in detail many of the technical issues required by the discretization and by the numerics to obtain an efficient iterative algorithm that scales to many hundreds if not thousands of processors. 

While this work is a significant advance in theoretical many-body physics, it is not clear that this phenomena can be observed directly in solid-state systems.  The problem there is that most solid state systems where Bloch oscillations can be observed have higher energy bands present.  If the field becomes large enough to induce tunneling between the bands, then the theory we have derived will not apply, because we have neglected Zener tunneling.  This implies the observation of beats may be difficult to see in a solid-state system.  Furthermore, the fields are so large here that the Bloch frequencies are enormous, with periods probably lying in the femtosecond range.  It is a challenge to find experimental techniques to measure such rapid oscillations.  

One system that might be appropriate for this experiment, however, is ultracold atoms placed in optical lattices.  If we consider a mixture of two species where one of the atomic species is much more massive than the other, and if the delocalized fermions are in a band that is well separated from the higher-energy bands, then by detuning the optical lattice, so that it appears as a static lattice in a moving frame (via the Doppler effect), one can ``pull'' the lattice through the atomic cloud, in direct analogy to the Hamiltonian gauge calculations presented here. Unfortunately there is no direct way measure the current, but it could be reconstructed, in principle, from a time-of-flight measurement which determined the distribution of atoms through the Brillouin zone.

In future work, we plan to examine the steady-state limit directly and how the distribution functions transiently evolve after the field is turned on (the latter being important for direct comparison to time-of-flight measurements in ultracold atoms).

\acknowledgments  
We acknowledge useful conversations with V. Turkowski, V. Zlati\'c,
J. Serene and A. Joura.  This work is supported by the N. S. F. under grants
numbered DMR-0210717 and DMR-0705266, and by the O. N. R. under grant number N000140510078.
Supercomputer time was provided on the ERDC XT3 under a CAP phase II 
project, on the ARSC Midnight SUN OPTERON under a CAP phase II project, and on the NASA Columbia SGI ALTIX under a National Leadership Computing System grant.

\end{document}